\documentstyle[preprint,prl,aps]{revtex}
\input epsf.tex
\def\eqn{\begin{equation}}
\def\endeqn{\end{equation}}
\def\eqna{\begin{eqnarray}}
\def\endeqna{\end{eqnarray}}
\begin{document}
\draft
\preprint{CLNS97/1463,~~NWU970201,~~hep-ph/9702218}
\title
{
Radiative Corrections to the Muonium Hyperfine Structure.  \\
II.  The $\alpha (Z\alpha)^2$ Correction }
\author{ M. Nio\thanks{electronic address: makiko@phys.nara-wu.ac.jp} }
\address{ Graduate School of Human Culture,
Nara Women's University, Nara, Japan 630 }

\author{T. Kinoshita\thanks{electronic address: tk@mail.lns.cornell.edu} }
\address{ Newman Laboratory of Nuclear Studies,
Cornell University, Ithaca, NY 14853 }
\date{\today}
\maketitle

\begin{abstract}
This is the second of a series of papers on the radiative corrections
of order $\alpha^2 (Z\alpha)$, $\alpha (Z\alpha )^2$, and various
logarithmic terms of order $\alpha^4$,
to the hyperfine structure of the muonium ground state.
This paper deals with  
the $\alpha (Z\alpha)^2$ correction.
Based on the NRQED bound state theory, we isolated  
the  term of order  $\alpha(Z\alpha)^2$ exactly. 
Our result $+16.904~2~(11) \alpha(Z\alpha)^2 E_F / \pi$ 
for the non-logarithmic part 
is consistent with the $\alpha (Z\alpha )^2$ part of 
Sapirstein's  calculation and the recent result of Pachucki,
and reduces the numerical uncertainty in the $\alpha (Z\alpha)^2$ term 
by  two orders of magnitude.  
\end{abstract}


\vspace{5ex}

\def\logtLg{\ln \biggl ({2\Lambda m \over \gamma^2}\biggr) }
\def\logstLg{\ln^2 \biggl ({2\Lambda m \over \gamma^2} \biggr ) }
\def\logstLl{\ln^2 \biggl ({2\Lambda m \over \lambda^2} \biggr ) }
\def\logLl{\ln \biggl ({\Lambda  \over \lambda} \biggr ) }
\def\logsLl{\ln^2 \biggl ({\Lambda  \over \lambda} \biggr ) }
\def\logtLml{\ln \biggl ({2\Lambda m \over \lambda^2} \biggr ) }
\def\logl2L{\ln \biggl ({\lambda  \over 2\Lambda} \biggr ) }
\def\logtml{\ln \biggl ({2m \over \lambda} \biggr ) }
\def\logt3{\ln{2 \over 3} }
\def\logml{\ln \biggl ({m \over \lambda} \biggr ) }
\def\logsml{\ln^2 \biggl ({m \over \lambda} \biggr ) }
\def\logl2L{\ln \biggl ({\lambda \over 2\Lambda} \biggr ) }
\def\logm2L{\ln \biggl ({m \over 2\Lambda} \biggr ) }
\def\loglg{\ln \biggl ({\lambda \over \gamma} \biggr ) }
\def\logLg{\ln \biggl ({\Lambda \over \gamma} \biggr ) }
\def\logmg{\ln \biggl ({m \over \gamma} \biggr ) } 
\def\logsmg{\ln^2 \biggl ({m \over \gamma} \biggr ) } 
\def\logmgs{\ln \biggl ({m^2 \over \gamma^2} \biggr ) } 
\def\logsLm{\ln^2 \biggl ({\Lambda \over m} \biggr ) } 
\def\logLm{\ln \biggl ({\Lambda \over m} \biggr ) }

\section{Introduction}
\label{sec:i}

Now that the $\alpha^2(Z\alpha)$ correction  to the hyperfine splitting
of the ground state muonium is known very accurately because of  recent 
works \cite{KN0,ES1,KN1},
the previously-calculated $\alpha(Z\alpha)^2$  term 
has become one of the main sources of the remaining theoretical uncertainty. 
Improvement of this error is urgently needed
in view of the new muonium hyperfine measurement in progress \cite{hughes}.
Here $Z$ is the $``$atomic" number of the nucleon. 
Of course $Z=1$ for
the muonium, but it is kept to indicate the bound state
origin of the correction terms.
In this paper, we present 
the $\alpha(Z\alpha)^2$ radiative correction 
evaluated
in the NRQED bound state formulation \cite{CL}.

The previous evaluation of this term, including recent 
modifications \cite{KN1,sapirstein2}, gives 
\eqna
\Delta \nu [\alpha(Z\alpha)^2] & =&  E_F {\alpha(Z\alpha)^2 \over \pi}
                \biggl [ -{8 \over 3} \ln^2(Z\alpha)^{-1} 
\nonumber \\
          &&
+ \left( - {8\over3}   \ln4 + {37\over36} + {8\over 15}
                       \right ) \ln (Z\alpha)^{-1} 
              +  14.88~(29)~\biggr ]~~~,
\label{oldBS}
\endeqna
where  the Fermi frequency $E_F$ is defined by \cite{fermi}
\eqn
  E_F={16 \over 3}\alpha^2 c R_{\infty} {{m} \over {M}}
  \left [ 1+{m \over M} \right ]^{-3},     \label{EF}
\endeqn
$R_{\infty}$ is the Rydberg constant for infinite nuclear mass,
and $m$ and $M$ are the electron and muon masses, respectively.
The presence of $\ln(Z\alpha)$ factors is 
due to the near infrared (IR) structure of the 
loosely bound state. 
The coefficients of $\ln^2(Z\alpha)^{-1}$ 
and $\ln(Z\alpha)^{-1}$
terms were obtained
by Layzer \cite{layzer} and Zwanziger \cite{zwanziger2} independently.
Brodsky and Erickson \cite{BE} confirmed these logarithmic terms and gave 
the leading contribution of the 
%
%
%
%
non-logarithmic term. 
Sapirstein reported the numerical evaluation of the  non-logarithmic 
constant due to the radiative photon   \cite{sapirstein}. 
For convenience's sake, we refer to this non-logarithmic constant of the
$\alpha(Z\alpha)^2$ correction as the BES term.

To compute the BES term, Sapirstein started from
the relativistic bound state 
formalism and evaluated the entire $\alpha(Z\alpha)^2$ term numerically.
In his approach, only the double logarithmic term  was confirmed by varying
the $``$atomic"  number $Z$.  Since the logarithmic term is a consequence
of the near IR singularity of the bound state, 
the convergence of 
%
%
%
%
%
%
numerical 
%
%
%
%
%
integration worsens in the 
%
%
%
%
%
%
region of  
small momentum.
The uncertainty in the BES term  comes mainly from 
this difficulty in  the numerical integration. 
Note also that his result  contains terms of higher orders in $Z\alpha$. 


Our calculation of the BES term starts from 
the NRQED formalism proposed by Caswell and Lepage \cite{CL}. 
This approach enables us  to isolate the $\alpha(Z\alpha)^2$ 
term without being tangled up with higher order terms in $Z\alpha $: 
All these terms arise from different parts of the NRQED Hamiltonian.  
The leading logarithmic contribution is analytically separated.    
The small photon mass $\lambda$ is used in our approach but 
the $\lambda$ dependence can be easily identified 
and analytically subtracted in the numerical 
%
%
%
%
%
evaluation
of 
each diagram. 
This is important for reducing the computational error of the BES term.

Asides from the factor $(\alpha(Z\alpha)^2  / \pi) E_F $
our results for the vacuum polarization contribution and 
the radiative photon contribution to the BES term are
\eqn
\Delta E_{\rm BESvp}= -0.218~567~\cdots~,
\label{myBESvp}
\endeqn
and
\eqn
\Delta E_{\rm BESph}= 17.122~7~(11)~,
\label{myBES}
\endeqn
respectively.
Our final result for the total  $\alpha(Z\alpha)^2$ correction is
\eqna
\Delta \nu [\alpha(Z\alpha)^2] & =&  E_F {\alpha(Z\alpha)^2 \over \pi}
                \biggl [ -{8 \over 3} \ln^2(Z\alpha)^{-1} 
\nonumber \\
          &&
+ \left( - {8\over3}   \ln4 + {37\over36} + {8\over 15}
                       \right ) \ln (Z\alpha)^{-1} 
              +  16.904~2~(11)~\biggr ]~~~.
\label{newBS}
\endeqna

The BES term 
has also been evaluated  recently by Pachucki \cite{pachucki1} using the method
he developed for the $\alpha(Z\alpha)^6 m $ Lamb shift 
correction \cite{pachucki2}.
His result for the radiative photon contribution is 
\eqn
\Delta E_{\rm BESph}= 17.122~.
\label{pacBES}
\endeqn
Although this is in good agreement with our result (\ref{myBES}), there remains
some disagreement in the details.  This will be discussed in Sec. VI.

In Sec. II, we briefly describe our approach, 
%
%
%
%
%
%
namely the NRQED method, to the
bound state problem.  A more detailed prescription of NRQED is 
found in Ref. \cite{KN1}.  
In Sec. III, the well-known Breit $(Z\alpha)^2$ correction \cite{breit} 
is rederived from NRQED.
This calculation, multiplied by an appropriate $``$renormalization" factor,
actually provides a part of the $\alpha(Z\alpha)^2$ term.
It also serves as a prototype of  the calculation of the entire 
$\alpha(Z\alpha)^2$ correction.
In Sec. IV  we derive the $\alpha (Z\alpha )^2$ contribution arising 
>from the vacuum polarization
insertion.  We uncovered a mistake in the previous calculation in 
Ref. \cite{BE} as was mentioned in our previous paper 
%
%
%
%
%
%
\cite{KN1}.
The detail of the $\alpha(Z\alpha)^2$
correction due to the radiative photon on the electron line
is described in Sec. V.
Sec. VI is devoted to the discussion of our result.


\section{Outline of the NRQED method}

%
%
%
%
%
%
The NRQED is a theory with a finite UV cut-off, which is completely equivalent 
to QED when it is applied to low-energy
systems with  typical momenta less than the UV cut-off $\Lambda$.
The NRQED Lagrangian consists of operators which satisfy the
%
%
%
%
%
%
same symmetries as the QED operators except that they
satisfy Galileian invariance,
although the final observable results of calculation
are Lorentz invariant.
Fermions are represented not by Dirac spinors but by Pauli spinors.
The NRQED Lagrangian  can be divided into two parts: 
The main part $L_{\rm main}$ 
consisting of  fermion bilinear operators
%
%
%
%
%
multiplied by up to two photon operators 
or pure photon terms
and the contact interaction part $L_{\rm contact}$
involving four or more fermions.
Both parts of the Lagrangian are determined by the following simple rule: 
The operators which appear in its Lagrangian
and their coefficients are chosen so that  
any $scattering$ amplitude calculated in the  NRQED
coincides with the corresponding $scattering$ amplitude
of the original QED at some given momentum scale,
e.g., at the threshold of the external on-shell particles.
This matching condition is applied order by order to the expansion 
in the coupling
constant $\alpha$ and velocity $v$ of the external fermion.
The Coulomb gauge is used in the NRQED, while  
the Feynman gauge is more convenient to compute the QED scattering diagrams.
Readers interested in  NRQED may refer to
Refs. \cite{CL,KL,nrqcd,positronium,labelle}.
The precise description of the NRQED Hamiltonian can be found in 
Ref.\cite{KN1}. 
After determining all operators and their coefficients to the
desired order of  velocity $v$ of the electron and the coupling constant
$\alpha$,
we evaluate the energy shift, etc. using the Rayleigh-Schr\"{o}dinger
perturbation theory, choosing as the unperturbed system the
exact solution of the nonrelativistic Schr\"{o}dinger Coulomb system.
   
The main part of the NRQED Hamiltonian $H_{\rm main}^{\Lambda}$ 
needed  to compute 
the $\alpha(Z\alpha)^2$ correction  terms is  of the form 
\eqna
H_{\rm main}^{\Lambda}  =  \psi^{\dagger}(\vec{p}\,')
        \biggl [ && { {\vec{p}\,^2} \over 2m } + eA^0
           - { (\vec{p}\,^2)^2 \over 8 m^3 }
           -{e \over 2m}  (\vec{p}\,'+\vec{p})\cdot\vec{A}
\nonumber \\
    &- & {ie \over 2m} c_F\vec{\sigma}
      \cdot (\vec{q}\times\vec{A})
           - {e \over 8m^2} c_D\vec{q}\,^2 A^0
\nonumber \\
&-& { i e^2 \over 4m^2} c_S \vec{\sigma}\cdot
             (\vec{q_2}\times \vec{A}(q_1) ) A^0(q_2)
        + (  q_1 \leftrightarrow q_2 ) 
\nonumber \\
 &+& {ie \over 8m^3} c_W(\vec{p}\,'^2+\vec{p}\,^2)\vec{\sigma}
                   \cdot (\vec{q}\times\vec{A})
     + {ie \over 8m^3} c_{q^2}\vec{q}\,^2\vec{\sigma}
                   \cdot (\vec{q}\times\vec{A})
\nonumber \\
    &+& {ie \over 8m^3} c_{p'p}
               \{ \vec{p}\cdot(\vec{q}\times\vec{A})
               (\vec{\sigma}\cdot\vec{p}\,')
                + \vec{p}\,'\cdot(\vec{q}\times\vec{A})
                     (\vec{\sigma}\cdot\vec{p}) \}~~~ \biggr ] \psi(\vec{p})
\nonumber \\
  &+&c_{\rm vp}A^i(\vec{q}){ \vec{q}\,^4 \over m^2 }
       A^j(\vec{q})(\delta^{ij}-{q^iq^j \over \vec{q}\,^2})
\nonumber \\
  &+&c_{\rm vp}A^0(\vec{q}){ -\vec{q}\,^4 \over m^2 }
     A^0(\vec{q})~,
\label{H_main}
\endeqna
where 
$\vec{p}$ and $\vec{p}\,'$  are  incoming and outgoing   
electron momenta, respectively, and $\vec{q}$ is incoming photon momentum.
\footnote{ The term $\vec{q}_1 \times \vec{A}(q_1)$ in  Eqs. (53) and (55) 
of  Ref.\cite{KN1}  must be replaced by  $\vec{q}_2 \times \vec{A}(q_1)$. }
The superscript $\Lambda$ in $H_{\rm main}^{\Lambda}$ indicates that
the theory is regularized by the UV cut-off $\Lambda $ for the radiative photon.
$\Lambda$  may be less than $m$.
The $``$renormalization" coefficients in (\ref{H_main}) are
\eqna
c_F& =&1+ a_e~,
\nonumber \\
c_D&=&1 + { \alpha \over \pi } {8 \over 3}
     \biggl [ \ln \left({m \over 2\Lambda }\right)
-{3 \over 8}+{5 \over 6}\biggr ]
                 + 2 a_e~,
\nonumber \\
c_S&=& 1+ 2a_e ~,
\nonumber \\
c_W&=& 1~,
\nonumber \\
c_{q^2}&=& {\alpha\over\pi} {4 \over 3}\biggl [
\ln\left({m \over 2\Lambda}\right)
- { 3 \over 8 } + { 5 \over 6}  +{1 \over 4}\biggr ]  + {a_e \over 2} ~,
\nonumber \\
c_{p'p}&=& a_e~,
\nonumber \\
c_{\rm vp}&=& {\alpha \over 15\pi } ~.
\label{Rconsts}
\endeqna
The first term in (\ref{H_main}) is the nonrelativistic kinetic energy term. 
The rest 
are named successively as Coulomb, 
$p^4$ relativistic kinetic energy,
dipole coupling, 
Fermi, Darwin , seagull, $W$-(wave function) derivative Fermi, $q^2$ derivative
Fermi, and $p'p$ coupling, respectively.
The last two terms bilinear in photon operators are introduced to represent the
vacuum polarization insertion in the transverse and Coulomb photon
propagator, respectively.

%
%
%
%
%
%
For the muon line, only the Coulomb and Fermi terms  
are needed 
for the calculation of the $\alpha(Z\alpha)^2$ correction.
They are obtained by replacing  $e$ and $m$ of the electron interaction terms
by $-Ze$ and $M$, respectively \footnote{We use the convention that 
the electron charge is $e$ and the positive muon charge is  $-Ze$.} . 

As for the contact part of the NRQED Hamiltonian $H_{\rm contact}^{\Lambda}$,  
only   the spin-flip type  is needed: 
\eqn
 H_{\rm contact}^{\Lambda} = 
 -d_1 {1 \over mM } (\psi^{\dagger} \vec{\sigma }\psi)
 \cdot (\chi^{\dagger} \vec{\sigma }\chi),
\label{H_contact}
\endeqn
where $\chi$ is the Pauli spinor for the positive muon and $d_1$ can be written as
\eqna
-d_1&=& \alpha(Z\alpha)^2 \pi {2 \over 3} 
      \biggl[ \ln 2 - {13 \over 4 }+{3 \over 4} \biggr ] 
\nonumber
\\
&+& \alpha(Z\alpha)^3 {2\over3}
\biggr[ d^{(3)}\sqrt{{\Lambda m \over ((Z\alpha)m)^2} }   
     +   d^{(2)} \logsLm + d^{(1)}\logLm + d^{(0)}
\nonumber \\
    & +&  d^{(4)}{\Lambda ' \over (Z\alpha)m }
     +  d^{(5)}\ln\left({\Lambda '  \over m} \right)  
                                   \biggr]. 
\label{d_1BS}
\endeqna
Here  $\Lambda'$ is the UV cutoff of the
transverse exchanged photon momentum. 
Note that, when the
$\alpha(Z\alpha)^2$ part of Eq. (\ref{d_1BS}) is evaluated in the
first order perturbation theory, it gives the 
$\alpha(Z\alpha)$ correction to the hyperfine splitting of the
ground state muonium calculated by 
Kroll and  Pollock, and Karplus, Klein, and J. Schwinger
\cite{KPoriginal}.  
For brevity, let us call the $\alpha(Z\alpha)$ correction as
the KP correction.
The coefficients $d^{(0)}, \cdots , d^{(5)}$ of the $\alpha(Z\alpha)^3$ part
of Eq. (\ref{d_1BS})  are pure numbers.  These are the quantities 
that we want to calculate in this paper.  

The  Green function $\tilde{G}_0(\vec{p},\vec{q};E)$
appearing in this calculation is known in an exact closed form
for the nonrelativistic Coulomb potential \cite{schwinger}. 
For an arbitrary energy $E$ 
in the complex $E$ plane this function takes the form:
\eqna
\tilde{G}_0(\vec{p},\vec{q};E)
&=& - { 2m \over \vec{p}\,^2 + p_0 ^2 } (2\pi)^3\delta^3(\vec{p}-\vec{q})
\nonumber \\
&&+  { -2m \over \vec{p}\,^2 + p_0 ^2 }
  { -4\pi Z\alpha  \over |\vec{p} -\vec{q}|^2 }
  { -2m \over \vec{q}\,^2 + p_0 ^2 }
\nonumber \\
&&- { 64\pi   \over \gamma^4 (Z\alpha)}
\tilde{R}(\vec{p},\vec{q};E) ~~,
\label{green1}
\endeqna
where
\eqn
p_0^2=-2mE,~~~~~~ \nu={\gamma \over p_0 }~,
\endeqn
and
\eqna
\tilde{R}(\vec{p},\vec{q};E)& = &
{ \gamma^7 p_0 \over (\vec{p}\,^2 +p_0^2)(\vec{q}\,^2+p_0^2) }
\nonumber \\ && \times
\int_0^1 d\rho{ \rho^{-\nu} \over 4\rho|\vec{p}-\vec{q}|^2p_0^2
            + (1-\rho)^2(\vec{p}\,^2+p_0^2)(\vec{q}\,^2+p_0^2)} ~~.
\label{green0}
\endeqna
The first, second, and third terms of (\ref{green1}) correspond to zero, 
one, and two or more Coulomb-photon exchanges. 
For $E=E_0-k$, where $E_0$ is the ground state energy and $k$ is the energy of
a radiative photon,
$p_0^2$ may be written as 
\eqn
p_0^2 = -2m(E_0 - k) = \gamma^2+ 2mk~.
\endeqn
We calculate the $(Z\alpha)^2$ and $\alpha(Z\alpha)^2$ corrections
in the subsequent sections
using the NRQED Hamiltonian and the Green function given above.

\section{The $ (Z\alpha)^2 $  Correction}


In this section, we rederive the well-known Breit $(Z\alpha)^2$ relativistic 
correction \cite{breit} from the NRQED in order to illustrate how it works,
particularly, how the contact term $H_{\rm contact}$ is constructed. 
The first computation of the Breit term in the framework of the NRQED was 
carried out in Ref.\cite{CL}.
In the previous paper \cite{KN1}, we have shown that  
both $\alpha(Z\alpha)$ and $\alpha^2(Z\alpha)$
corrections come from the first order perturbation theory of 
the $H_{\rm contact}^{\Lambda}$ which represents the difference
between  the QED scattering amplitude  and 
the NRQED scattering amplitude.  
In contrast, the contact term calculated here
is derived entirely  from the NRQED scattering amplitudes.  
Their presence is crucial to cancel the UV divergence occurring 
in the bound state calculation of the operators in 
the $H_{\rm main}^{\Lambda}$.
More importantly, the calculation carried out in this section  immediately
yields  a part of the $\alpha(Z\alpha)^2$  correction 
if one takes account of appropriate 
$``$renormalization" constants  of the potentials.  
We must make sure that this calculation is consistent with the calculation
of the  other parts of 
the $\alpha(Z\alpha)^2$ correction.  This is why
we want  to rederive the Breit correction  by the NRQED method,
not by other bound state formalism.

\subsection{Diagram Selection}

The first task is to identify the potentials contributing
to the Breit $(Z\alpha)^2$ correction. 
This can be achieved using the order of magnitude estimate of 
various operators appearing in
the $H_{\rm main}^{\Lambda} $ of Eq. (\ref{H_main}) \cite{nrqcd} 
\eqna
&& < \vec{\partial}> \sim  m(v/c), ~~~
< \partial_t> \sim  m(v/c)^{2},~~~
< eA^0>  \sim  m(v/c)^{2},
 \nonumber \\
&&<e\vec{A}>  \sim  m(v/c)^{3}, ~~
<e\vec{E}>  \sim m^{2}(v/c)^{3}, ~~
<e\vec{B}>  \sim m^{2}(v/c)^{4},   
\label{order}
\endeqna
where $(v/c) \sim (Z\alpha)  $.
%
%
%
%
%
%
For example, for the Coulomb potential between the electron and the muon  
\eqn
V_C(\vec{p}\,',\vec{p}) \equiv  {- Z e^2  \over \vec{k}\,^2 + \lambda^2 },
\label{V_C}
\endeqn
where $\vec{k}=\vec{p}\,'-\vec{p}$, 
%
%
%
%
%
%
we have $<V_C > /m \sim (v/c)^2 \sim (Z\alpha)^2 $.
We  will set the photon mass $\lambda$ to  zero in 
the bound state calculation. 

Since the Fermi potential 
\eqn
V_F(\vec{p}\,',\vec{p}) \equiv 
  \frac{ -ie\vec{k} \times\vec{\sigma_{e}} }{2 m}
          \cdot
          \frac{iZe(-\vec{k})\times \vec{\sigma_{\mu}}}{2M}
          {-1 \over \vec{k}\,^2 + \lambda^2}
\label{V_F}
\endeqn
%
%
%
%
%
%
has an expectation value of order $(Z\alpha)^4 (m/M)m$,
one source of the Breit  $(Z\alpha)^{2}E_F$ correction   
is  the first order perturbation with the order 
$(Z\alpha)^{6} (m/M) m$ potentials. 
All potentials of this type must have  spin flipping property
in order to contribute to the hyperfine structure. 
%
%
%
%
%
%
One of these potentials is the $W$-derivative Fermi term 
in Eq. (\ref{H_main}) which yields the potential
\eqna
 V_{W}(\vec{p}\,',\vec{p})& \equiv & \frac{ -ie(\vec{p}\,^{2}+\vec{p}\,'^{2})\vec{k}
          \times\vec{\sigma_{e}} }{8 m^{3}}
          \cdot
          \frac{iZe(-\vec{k})\times \vec{\sigma_{\mu}}}{2M}
          {-1 \over \vec{k}\,^2 +\lambda^2}~,
\label{V_W}
\nonumber \\
 \langle V_{W} \rangle   & \sim &\frac{(Z\alpha)\gamma^{5}}{m^{3}M}
      \sim (Z\alpha)^{6}{m \over M} m~.
\endeqna
Another contribution comes from  the seagull term: 
\eqna
 V_{S}(\vec{p}\,',\vec{p},\vec{q})&\equiv& \frac{ie^{2} \vec{q}\times \vec{\sigma_{e}} }{4m^{2}} 
          \cdot
  \frac{iZe(-\vec{k})\times\vec{\sigma}_{\mu}}{2M} \frac{-1}{\vec{k}\,^2+ \lambda^2}
          (-Ze) \frac{1}{\vec{q}\,^2 + \lambda^2}~,
\label{V_S}
\nonumber \\
 \langle V_S \rangle  & \sim & \frac{(Z\alpha)^{2}\gamma^{4}}{m^{2}M}
      \sim (Z\alpha)^{6}\frac{m}{M} m~,
\endeqna
where  $\vec{k}$ and $\vec{q}$  are the transverse and Coulomb
photon momenta, respectively.

One must also consider  possible contributions from higher order perturbation
theory. The second order perturbation term has the form
\eqna
  \delta E & = &  \langle  \psi_{n=1}| V 
          (\tilde{G}_{0}-\frac{ |\psi_{n=1} 
\rangle \langle \psi_{n=1}|} {E-E_{n=1}^{0}})
                   V | \psi_{n=1}\rangle _{E=E_{n=1}^{0}} 
\nonumber \\
           & = & \sum_{k \neq 1} 
           \frac { \langle  \psi_{n=1}| V |\psi_{k}\rangle
                   \langle  \psi_{k}| V |\psi_{n=1}\rangle  }
           {E_{n=1}^{0}-E_{k}^{0}}~.
\label{perturb}
\endeqna
Since the denominator $ E_n^0-E_k^0 $ is of order $ (Z\alpha)^{2} m $, 
one potential must be of order $(Z\alpha)^{4} (m/M) m$ 
while the other is of order $(Z\alpha)^{4} m $ in order that $\delta E$ 
contributes to the Breit correction:
\eqn
\delta E  \sim  \frac{ (Z\alpha)^{4} (m/M)m  (Z\alpha)^{4} m }
                                   { (Z\alpha)^{2} m }
           \sim (Z\alpha)^{6} \frac{m}{M}m ~. 
\endeqn
This can be realized only if one is  the Fermi potential
and the other is an order $(Z\alpha)^{4} m $  spin-non-flip
potential. We find  two candidates for the latter: the relativistic
kinetic energy term
%
%
%
%
%
%
\eqna
 V_{K}(\vec{p}\,', \vec{p})
&\equiv& 
 V_{K}(\vec{p})
          (2\pi)^{3}\delta^{3}(\vec{p}-\vec{p}\,')
\nonumber \\
&\equiv& - { (\vec{p}\,^{2})^{2} \over 8 m^{3}}
          (2\pi)^{3}\delta^{3}(\vec{p}-\vec{p}\,')~,
\label{V_K}
\nonumber \\
 \langle V_K \rangle &\sim & {\gamma^{4}\over m^{3}}
     \sim  (Z\alpha)^{4} m ~,
\endeqna
and the Darwin term
\eqna
 V_{D}(\vec{p}\,',\vec{p})&\equiv&  \frac{-e \vec{k}^{2}}{8 m^{2}}
         (-Ze) \frac{1}{\vec{k}\,^2 + \lambda^2}~,
\label{V_D}
\nonumber \\
\langle V_D \rangle & \sim &\frac{(Z\alpha)\gamma^{3}}{m^{2}}
     \sim  (Z\alpha)^{4} m  ~.
\endeqna
The third  and higher order perturbation terms   
do not  contribute to the Breit correction.

\subsection{Determination of the NRQED Contact Terms }

We have identified the NRQED potentials necessary for the calculation
of the Breit correction, namely, the Fermi, derivative Fermi, 
seagull, relativistic kinetic energy, and Darwin terms.
The next step is the determination of  NRQED 
contact terms corresponding to  these potentials.
The  contact terms of spin-non-flip type contribute to the 
hyperfine splitting  calculation
only through the second- or higher-order perturbation terms,
analogous to the case of the Darwin or relativistic kinetic energy potentials. 
However, 
there is no $(Z\alpha)^4 m $ contact term  
of the spin-non-flip type.
The lowest order spin-non-flip contact term   
is of order   $\alpha(Z\alpha)^5 m$, which gives the relativistic binding
correction to the Lamb shift.  
Therefore, we have only to consider the NRQED contact terms
of spin-flip type 
\eqn
\delta H= -d_1 {1 \over mM }( \psi^{\dagger} \vec{\sigma_e }\psi )\cdot
(\chi^{\dagger} \vec{\sigma_{\mu} }\chi ).
\endeqn

In the following the muon is treated as an external static photon source 
since the infinite muon mass limit is taken.  
However, the method of the contact term determination for 
the dynamical muon case is essentially the same as  the static case 
and the latter can be readily adapted to the former.

\begin{figure}
\centerline{\epsfbox{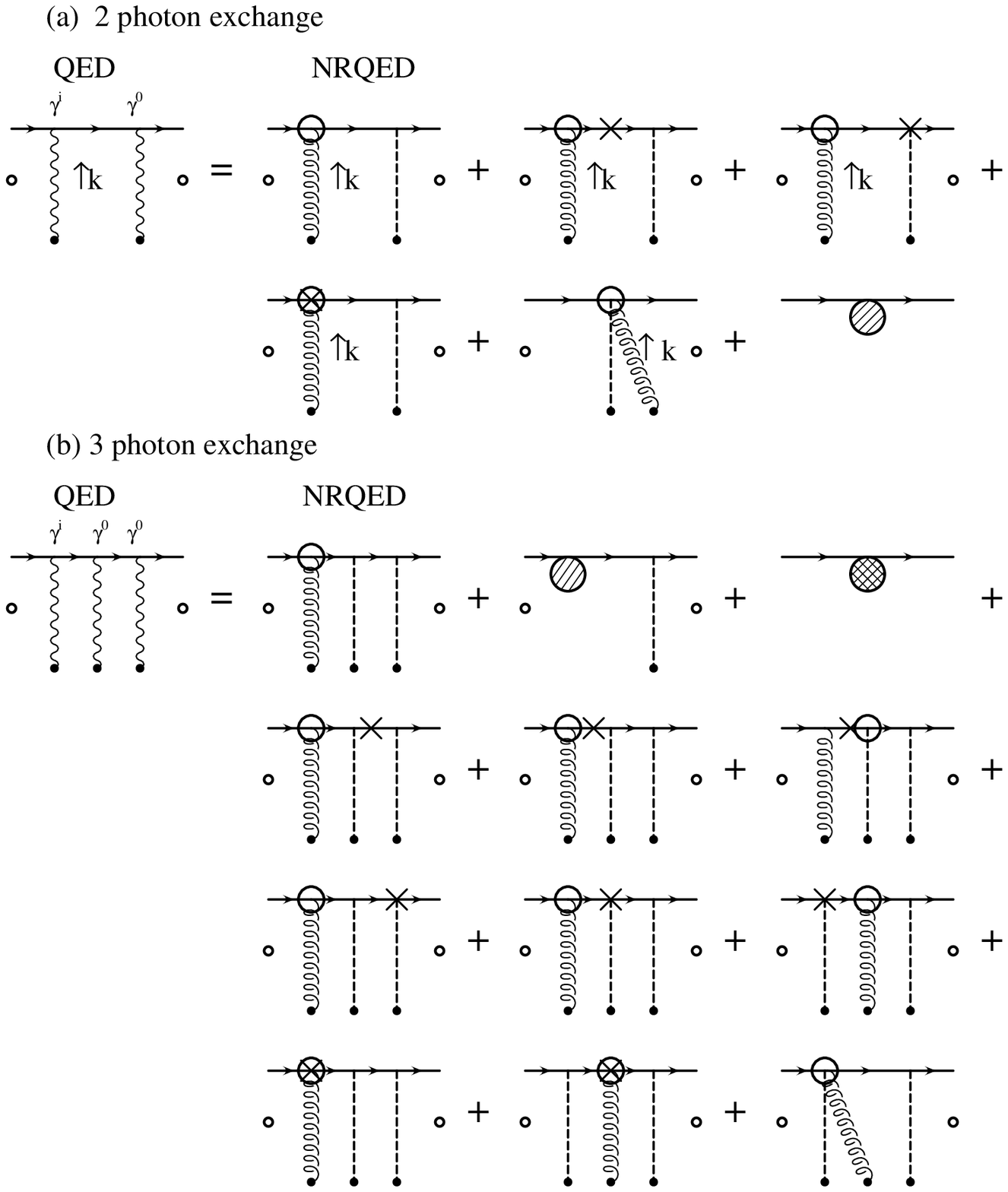}} 
\caption{Comparison of the QED and NRQED scattering amplitudes. 
The small circles on the edges of each diagram indicate that 
a diagram is  evaluated on-the-mass-shell
and at the threshold external fermions. 
The shaded circle in (a) is a contact  term.
The doubly-shaded circle in (b) is a contact term, too.
A wavy line is  covariant photon in the Feynman gauge,
a carly and dashed line are  transvers and  Coulomb photons,
respectively, in the Coulomb gauge. 
The muon is treated as the static external source and indicated by 
a dot.    
For NRQED electron vertices, consult  the NRQED Feynman rules
given in Ref. [3] . 
\label{Bhamil}}
\end{figure}

As was explained in  Ref.\cite{KN1}, the contact terms are determined
by comparison of QED $scattering$ amplitudes and NRQED $scattering$ amplitudes.
The lowest order diagrams contributing to the contact term  relevant to
the Breit correction
is the two-photon-exchange process between the electron
and muon, namely,  one-loop process.
In order to contribute to the Breit correction, at least 
one of the photons must be transverse.  
Exchange of two transverse photons results in a recoil type 
correction proportional to $(m/M)^2$ and is not of interest here.
That leaves diagrams with one transverse photon and 
one Coulomb photon.  
In QED, there is one diagram with $e\gamma^{i}$ and $e\gamma^{0}$
vertices. 
NRQED interaction terms which give, 
in the {\it bound state} calculation, 
higher-order contributions than the order 
we are interested in  are to be ignored in the {\it scattering}
amplitude comparison.
This reduces the relevant NRQED scattering amplitudes to the following five
combinations: a Coulomb potential
with a Fermi potential, 
a Fermi potential with a relativistic kinetic energy 
term and a Coulomb potential, a Fermi potential with 
a Darwin  potential, a Coulomb potential with a derivative 
Fermi potential, and a seagull potential. 
The first four are given by the second order perturbation 
theory of the NRQED Hamiltonian
and the fifth  is from the tree NRQED Hamiltonian.
These five together determine the contact terms represented
by the shaded circle in Fig. \ref{Bhamil}(a).
Only the $p^2 /2m$ part  of the NRQED Hamiltonian
is treated as unperturbed system for the scattering
perturbation theory. In other words, the Coulomb potential 
appears as one of perturbative potentials in this comparison.
The comparison are  shown in Fig. \ref{Bhamil}(a).

In this case the QED scattering amplitude at threshold is completely 
replicated by the NRQED scattering amplitude consisting of a Coulomb 
potential and a Fermi potential.  
Thus the contact term  must be chosen as  
{\it minus sign } times  the sum of the remaining four NRQED 
scattering amplitudes.

The one-loop contact terms determined in this comparison
are all linearly divergent and hence their values 
depend on how they are regularized.  
Although  gauge invariant regularization method is 
desirable, it is possible in this case to use
a simple momentum cut-off.
This is because, as we shall see in  Appendix A,
calculation of the $bound$ $state$ expectation value also leads 
to a divergent integral and must be regularized.
Even though the regularized bound state calculation 
is not gauge invariant,
the gauge invariance can be restored  if one chooses appropriate
contact terms. 
What is crucial is that the regularization method 
of the contact term is consistent with 
the bound state calculation \cite{pat}. 
We satisfy this requirement by introducing an UV cut-off $\Lambda$ 
in the momentum of the
transverse exchanged photon in both 
contact term calculation and bound state calculation.

The NRQED scattering diagrams can be easily written down using the
NRQED Feynman rule given  in Fig. 3  of Ref. \cite{KN1}.
Against the relativistic kinetic energy term this rule leads to 
the contact term of the form
\eqn
V_{1c-k}  = -2  \int^{\Lambda}
             \!\frac{d^{3}k}{(2\pi)^{3}}  V_F(0,\vec{k})
               \frac{-2m}{\vec{k}\,^2}
              V_K(\vec{k})
               \frac{-2m}{\vec{k}\,^2}
              V_C(\vec{k},0) ~.
\label{V_1c-k}
\endeqn
Here the factor two is to take account of the time-reversed diagram.
The contact terms against the Darwin term, derivative Fermi term, and seagull 
term can be similarly 
constructed and evaluated.
Explicit evaluation of (\ref{V_1c-k}) and other terms is carried out  
in Appendix A.

Next we consider the three photon exchange process, or
two-loop one,  shown in  Fig. \ref{Bhamil}(b).
This requires adding one more Coulomb  photon exchange potential 
to both QED and NRQED amplitudes of the two photon exchange type. 
Adding another type of potential of NRQED, such as the Darwin potential,  
is not necessary since it gives 
a power of $v^2 \sim (Z\alpha)^2 $ higher than what we are interested in.
However, it is necessary to consider new kinds of scattering amplitudes, 
which are the combinations
of the contact terms introduced in the two photon exchange
comparison and one Coulomb photon potential.
Again, the QED  (left-hand-side of Fig. \ref{Bhamil}(b)) contribution is 
equivalent to the first diagram of the NRQED contribution (on the 
right-hand-side of Fig. \ref{Bhamil}(b)).
Thus the contact terms are determined by the
rest of the NRQED scattering amplitudes. 
Processes with four or more photon  exchange
clearly have more explicit powers of
$Z\alpha$,  yielding  higher order contact terms. 
Thus we don't have to deal with them as far as we are interested 
only in the $(Z\alpha)^2$ Breit correction. 

Some of the two-loop contact terms, which have a Coulomb photon
exchange potential sandwiched between two other kinds of  potentials,  
are logarithmically divergent in both UV and IR regions 
and will kill UV divergences in the bound state calculation.
Others are linearly divergent in both UV and IR.  
These diagrams have the structure such that
one Coulomb photon potential is added to the edge of
one loop linear UV divergent diagrams proportional to
$\Lambda/m$ introduced in the 
two photon exchange comparison. 
The additional Coulomb photon on the edge yields the threshold
IR singularity  causing the term to be proportional to $m/\lambda$,
where $\lambda$ is  the photon mass introduced as the IR cut-off. 
The multiplication of the UV and IR divergences results in 
the form $\Lambda/\lambda$. 
The UV-IR singularity $\Lambda/\lambda$  appearing in these diagrams
is completely canceled out by the diagram which  consists of   
the one-loop contact term introduced  before 
and  the Coulomb photon potential connected by the free 
fermion propagators.    

Here we show one of the
two-loop contact terms against the relativistic kinetic energy term: 
\eqn
V_{2c-k(1)}  =  -2 \int^{\Lambda}\!\frac{d^{3}k}{(2\pi)^{3}}
                    \int\!\frac{d^{3}l}{(2\pi)^{3}}
         V_F(0,\vec{k})
          \frac{-2m}{ \vec{k}^{2}}
         V_C(\vec{k},\vec{l})
              \frac{-2m}{ \vec{l}\,^2}
             V_K(\vec{l})
              \frac{-2m}{ \vec{l}\,^2}
             V_C(\vec{l},0) ~.
\label{V_2c-k(1)}
\endeqn
Other contact terms related to the relativistic kinetic term in this order
are presented in the Appendix A.
The potential $V_{2c-k(1)}$ is logarithmically divergent.
Other terms are linearly divergent.

The contact term against the combination of the one-loop contact term
and the Coulomb potential, which we denote with the suffix $1loop^2$,
is  also linearly divergent and is given by
\eqn
V_{2c-k,1loop^2}  = 4 \int^{\Lambda}\!\frac{d^{3}k}{(2\pi)^{3}}  
                      \int\!\frac{d^{3}l}{(2\pi)^{3}}  
         V_F(0,\vec{k})
          \frac{-2m}{\vec{k}\,^2} 
          V_K(\vec{k})
          \frac{-2m}{\vec{k}\,^2} 
          V_C(\vec{k},0)
          \frac{-2m}{\vec{l}^2} 
          V_C(\vec{l},0) ~.
\label{V_2c-kloop}
\endeqn
Other two-loop contact terms involving the Darwin, 
derivative Fermi, and seagull terms are given
in  Appendix A.

\subsection{Summary of the $(Z\alpha)^2$ Correction}

We have prepared the non-radiative NRQED 
Hamiltonian, including the contact terms, 
up to the order $(Z\alpha)^6(m/M)m$:
\eqna
H_{{\rm NRQED}}^{\Lambda}
& = &\psi^{\dagger}~\biggl [~ \frac{\vec{p}\,^2}{2m} -\frac{Z\alpha}{r}
 + \frac{e}{2m}\vec{\sigma_{e}}\cdot \vec{B}
 -   { (\vec{p}\,^{2})^{2} \over 8 m^{3}}
 +   \frac{e \vec{\nabla}\cdot \vec{E}}{8 m^{2}}
\nonumber \\
&& + \frac{e \{ \vec{p}\,^{2},\vec{\sigma_{e}}\cdot\vec{B}\}}{8m^{3}}
 -  \frac{e^{2} \vec{\sigma}\cdot\vec{A}\times \vec{E}}{4m^{2}}~\biggr]\psi
\nonumber \\
& - & d_1 {1 \over mM } (\psi^{\dagger} \vec{\sigma_e }\psi)
 \cdot (\chi^{\dagger} \vec{\sigma_{\mu} }\chi)~.
\endeqna
The contact term coefficient $d_1$ is calculated in Appendix A. 
It is the sum of the contributions from the relativistic kinetic energy,
Darwin, derivative Fermi, and seagull terms:
\eqna
-d_1 &  = & (Z\alpha)^3 {2\over 3} \pi
\nonumber \\
& \biggl [ & -  2{\Lambda \over \gamma \pi} 
-2\ln\left({\Lambda \over \lambda}\right)
-4\ln2+6\ln3
\nonumber \\
&   & + {\Lambda \over \gamma \pi} 
+\ln\left({\Lambda \over \lambda}\right)
+2\ln2-3\ln3+{1 \over 2}
\nonumber \\
&   & +2{\Lambda \over \gamma \pi} +4\ln2-4\ln3
\nonumber \\
&   & - {\Lambda \over \gamma \pi} 
+\ln\left({\Lambda \over \lambda}\right)
 -2\ln 2+\ln 3- {1 \over 2} \biggr ]
\nonumber \\
& = & 0~.
\endeqna
The value of $d_1$ could vary for  different regularization methods. 
Although $d_1$ adds up to zero in our regularization method,
this does not mean that we do not need the contact term.
Finiteness and gauge invariance of the final answer are 
guaranteed  by the presence of this contact term in individual terms. 

Using  the non-relativistic Coulomb system as the unperturbed system, 
%
%
%
%
%
%
we can now calculate the binding effect, namely, 
the Breit hyperfine energy correction  in 
perturbation theory.
The results are summarized here as terms proportional to the 
Fermi energy $E_{F}$.
$\Delta E_{k}$, $\Delta E_{d}$, $\Delta E_{w}$, and $\Delta E_{s}$
%
%
%
%
%
%
are the contributions from the relativistic kinetic energy,
Darwin,  derivative Fermi, and seagull terms, respectively 
(see Figs. \ref{Kinetic}, \ref{Darwin}, \ref{Deri}, and \ref{SEAgull}):
\eqna
\Delta E_{k}
  & = &  E_{F}(Z\alpha)^{2} \biggl [-\frac{1}{2}-6\ln\left(\frac{2}{3}\right)
   +2\ln\left(\frac{\lambda}{\gamma}\right)\biggr]   ~,   \label{kinetic} \\
\Delta E_{d}
  & = &  E_{F}(Z\alpha)^{2}\biggl[3\ln\left(\frac{2}{3}\right)
        -\ln\left(\frac{\lambda}{\gamma}\right)\biggr] ~,
                                          \label{darwin}   \\
\Delta E_{w}
    & = &  E_{F}(Z\alpha)^{2}\biggl[\frac{5}{2}+4\ln\left(\frac{2}{3}\right)
\biggr] ~,
                                           \label{dfermi}  \\
\Delta E_{s}
    & = &  E_{F}(Z\alpha)^{2}\biggr[-\frac{1}{2}-\ln\left(\frac{2}{3}\right)
         -\ln\left(\frac{\lambda}{\gamma}\right)\biggl]~,    \label{seagull} 
\endeqna
where $\gamma=Z\alpha m$ is the typical momentum
scale of the muonium in the infinite muon mass limit, and 
$\lambda$ is the photon mass which is set to zero
at the end of the calculation. 
The sum of Eqs. (\ref{kinetic}), (\ref{darwin}), (\ref{dfermi}),
and (\ref{seagull}) gives 
\eqn
\Delta \nu [(Z\alpha)^2] = \frac {3}{2}  E_{F}(Z\alpha)^{2}~,
\endeqn
which is the well-known Breit correction.
Note that the  
IR singularities $\ln(\lambda/\gamma)$   cancel out
when all diagrams are summed up. 
In  Appendix A, we show details of derivation of these terms.

We have shown how the NRQED  bound state formalism works 
using the well-known 
Breit correction to the muonium hyperfine structure as an example. 
As we have seen, the contact term in NRQED plays 
the crucial role:  it describes the high energy behavior,
recovers the symmetry, such as Lorentz
symmetry and  gauge invariance, and 
kills the would-be IR- and UV-divergent quantities.

\section{The $\alpha (Z\alpha )^2$ Vacuum Polarization Contribution }

The order $\alpha$ radiative correction in the $\alpha(Z\alpha)^2$ term 
comes from two sources: One is from  the vacuum polarization 
insertion in one of 
the exchanged photons between the electron and muon
and the other is from  the  spanning  
photon on the electron line.
They are separately gauge invariant.  
In this section, we will deal with the contribution coming from the
vacuum polarization insertion.

The result of 
our numerical evaluation of the vacuum-polarization contribution was 
\eqn
\Delta \nu [\alpha(Z\alpha)^2]_{\rm vp}
 = {\alpha(Z\alpha)^2 \over \pi } E_F  \left[ 
                {8 \over 15} \logmg - 0.218~81~(29)   \right]~, 
\label{ourvp}
\endeqn
where the error comes from the numerical integration.
The error associated with the finite  photon mass, which was used as an
IR regulator, is of order $\lambda/m$, hence negligible for 
the case $\lambda/m  = 10^{-5}$.
Our result Eq. (\ref{ourvp}) 
disagreed with that obtained by Brodsky and Erickson \cite{BE}:
\eqna
\Delta \nu [\alpha(Z\alpha)^2]_{\rm vp}
& = &{\alpha(Z\alpha)^2 \over \pi } E_F  \left[ 
                {8 \over 15} \logmg -{8\over15}\ln2 
                +{214\over 225}-{2\over 3}   \right] 
\nonumber \\
             &= &{\alpha(Z\alpha)^2 \over \pi } E_F  \left[ 
                {8 \over 15} \logmg -0.085~234 \cdots  \right]~. 
\label{BEvp}
\endeqna
This term actually consists of two parts. One corresponds to
the vacuum polarization insertion in the Coulomb photon, 
and the other is  insertion in the transverse photon.
Our result for the first part agrees with the corresponding result of
Brodsky and Erickson. For the second part,
however, we found $-0.801~(4)$ instead of $-2/3$ in Eq. (\ref{BEvp}).
In an effort to  determine the cause of the discrepancy,
we have analytically evaluated 
the integral expressing  the vacuum
polarization contribution,  
given  by  Zwanziger in Ref. \cite{zwanziger}.
Our subsequent analytic work showed that the contribution for insertion in the
transverse photon is $ -4/5 $, in agreement with 
our numerical result. 
Using this corrected value, the numerical value of the non-logarithmic 
part of $\Delta \nu[\alpha(Z\alpha)^2]_{{\rm vp}}$ becomes 
\eqna
\Delta E_{\rm BESvp}&= &{\alpha(Z\alpha)^2\over \pi }E_F
\biggl[-{8\over 15}\ln2 + {214\over 225}-{4\over 5} \biggr ]
\nonumber \\
&= &{\alpha(Z\alpha)^2\over \pi }E_F ( -0.218~567~\cdots)~.
\label{BESvp}
\endeqna
Since we found   the error in the calculation
of Ref.\cite{BE} (in June 1994),
Sapirstein has also noticed it independently \cite{sapirstein3}, 
and Brodsky and Erickson have agreed with
the corrected value \cite{brodsky}.
We were also informed by Karshenboim \cite{karsh4} that
the same result was independently obtained by
Schneider, Greiner and Soff \cite{SGS}.

Let us briefly describe the NRQED evaluation of Eq.(\ref{ourvp}).
The terms relevant to the calculation of the NRQED Hamiltonian (\ref{H_main}) are 
the Coulomb, Fermi and two-photon  interaction terms. 
Combining these three interaction terms, we can construct the 
effective spin-flip and spin-non-flip potentials:
\eqn
V_{\rm tvp} \equiv { \pi (Z\alpha) \over mM } c_{\rm vp}
{ \vec{q}\,^2 \over m^2 }
(\psi^{\dagger}\vec{q}\times\vec{\sigma}_e\psi )\cdot
(\chi^{\dagger}\vec{q}\times\vec{\sigma}_{\mu}\chi)
{\vec{q}\,^2 \over (\vec{q}\,^2+\lambda^2)^2 }
\label{Vtvp}
\endeqn
and
\eqn
V_{\rm cvp} \equiv -{4\pi Z\alpha  \over m^2} c_{\rm vp}
(\psi^{\dagger}\psi) (\chi^{\dagger} \chi)
{\vec{q}\,^4 \over (\vec{q}\,^2+\lambda^2)^2 } ~,
\label{Vcvp}
\endeqn
respectively.
We  will construct the contact term of the NRQED Hamiltonian  
of Eq.(\ref{H_contact})
by comparing the QED and NRQED scattering amplitudes using the same recipe
as for the Breit correction discussed in Sec. II.
In this case both QED and NRQED scattering amplitudes contribute to the 
contact term. (See Fig. \ref{BSvp}).  
We evaluate the effect of these  potentials using the nonrelativistic
Rayleigh-Schr\"{o}dinger perturbation theory.

To see the UV and IR cancellation of the calculated result, 
however, it is more convenient to decompose the contribution 
of $H_{\rm contact}^{\Lambda}$ into several parts.  
The detail is described in Appendix B1.
In this treatment,
the bound state contribution comes from  the 
first-order perturbation of the spin-flip 
potential $V_{\rm tvp}$ of Eq. (\ref{Vtvp}) and from 
the second-order perturbation
involving  the spin-non-flip potential $V_{\rm cvp}$ of Eq. (\ref{Vcvp}) 
and the Fermi potential $V_F$.
The UV divergences in both bound state calculations are taken
care of by the corresponding contact terms derived from  the 
scattering amplitudes of these NRQED potentials. 
The sum of two contributions is found to be 
\eqn
\Delta E_{\rm NRQED vp}={\alpha(Z\alpha)^2 \over \pi }E_F {8\over 15} 
 \left [\loglg - 9\ln2 + {15 \over 2}\ln 3 - { 5\over 4}    \right]~.
\label{Enrqedvp}
\endeqn

The remainder is a part of the $H_{\rm contact}^{\Lambda}$ coming from the
QED scattering amplitude. 
The three-photon-exchange diagram with one vacuum-polarization
insertion contributes to this order. Since, at the two-photon-exchange
level, we have introduced the contact term in NRQED Hamiltonian 
representing the vacuum polarization effect of the $\alpha(Z\alpha)$ 
KP term, which is the $-$ 3/4 term on the first line of  Eq. (\ref{d_1BS}),
we must also take into account the contribution of 
this term  in evaluating the NRQED scattering amplitude. 
The NRQED scattering diagram
with the Kroll-Pollock (KP) contact term connected with the Coulomb 
potential by the free electron propagator should be subtracted
>from the QED scattering amplitudes. 
Their numerical value, for the photon mass $\lambda/m=10^{-5}$, is
\eqna
\Delta E_{\rm QED vp}
& =& {\alpha(Z\alpha)^2 \over \pi }E_F  (5.520~74~(29))
\nonumber \\
& =& {\alpha(Z\alpha)^2 \over \pi }E_F 
\biggl[ {8\over15}\logml-0.619~49~(29)\biggr] ~~.
\label{Eqedvp}
\endeqna
The sum of $\Delta E_{\rm NRQED vp}$ and $\Delta E_{\rm QED vp}$ gives 
the result of Eq. (\ref{ourvp}). 
The cancellation of the photon mass dependence between 
$\Delta E_{\rm NRQED vp}$ and $\Delta E_{\rm QED vp}$ has been 
checked numerically  using various values of $\lambda$.  
%
%
%
%
Of course Eq. (\ref{ourvp}) is superseded by the analytic
result which incorporates (\ref{BESvp}).

\section{The $\alpha (Z\alpha )^2$ Radiative Photon  Contribution }

More  challenging is the radiative  photon  
contribution to the $\alpha(Z\alpha)^2$ term.   
The sources of the $\alpha(Z\alpha)^2$ correction have been 
discussed in Ref.\cite{KN1}. 
Starting with the main part of the NRQED Hamiltonian (\ref{H_main}),
we can construct the contact term $ H_{\rm contact}^{\Lambda}$. Then  
we find four  types of contributions to the $\alpha(Z\alpha)^2$ 
correction in the bound state perturbation theory.  
The manifest cancellation of  UV and IR divergences  
occurs between the bound state calculation of the operators from 
the $H_{\rm main}^{\Lambda}$ and the corresponding part of the contact term 
derived from the same operators. 
For simplicity we omit the overall factor $(\alpha(Z\alpha)^2 /\pi) E_F$
in the following.
This means also that the anomalous magnetic moment $a_e$ stands for $1/2$.

One contribution, denoted $E_B$, 
where $B$ implies the bound state effect,
is the third order perturbation  with two dipole couplings
and one Fermi potential from the main part of the NRQED Hamiltonian,
$H_{\rm main}^{\Lambda}$. 
(See Fig. \ref{E_Bfig}.)
The radiative correction of order $\alpha$ 
comes from the presence of an intermediate virtual photon.  
For this calculation, we use the Coulomb Green's function 
in the presence of a photon with energy $k$  obtained
>from Eq. (\ref{green0}).
The UV cut-off $\Lambda$ is imposed on  the radiative transverse photon
momentum.
All UV divergent parts and some finite parts are analytically calculated.
Remaining finite parts are numerically evaluated using VEGAS \cite{lepage}.
The results are listed in Table \ref{E_Btable}. 
All relevant formulas are given in Appendix B2.

The second contribution  $E_S$ comes from the contact interaction
part of $H_{\rm contact}$ corresponding to the bound state calculation 
of $E_B$. 
The NRQED scattering diagrams  with two dipole couplings
and one Fermi potential  yield this contact interaction term in NRQED. 
(See Fig. \ref{E_Sfig}.)
The UV cut-off $\Lambda$ is also imposed on the radiative transverse 
photon momentum.
The calculation is entirely analytic.
The results are listed in Table \ref{E_Stable}.

The third type of  contribution $E_R$,  
where $R$ stands for the $``$R"enormalization constant,
is found in the first and second order perturbation terms with the
potentials which give the $(Z\alpha)^2$ correction. 
These potentials are found  using the analog of  the method of 
the Breit $(Z\alpha)^2$ correction. 
In this case,  the radiative correction of order $\alpha$ comes from the
$``$renormalization" coefficients of order $\alpha$ of these potentials.
The UV cut-off $\Lambda'$ is imposed on the exchanged transverse photon
momentum. This $\Lambda'$  cancels out when the bound state calculation 
is completed.
%
%
%
%
%
%
Some of these contributions are already calculated in Sec. III concerning
the Breit correction.
For the contributions from the relativistic kinetic energy, 
Darwin, and seagull term, we only need to multiply 
Eqs. (\ref{kinetic}), (\ref{darwin}), and (\ref{seagull}) with  
the appropriate $``$renormalization" constants given in Eq. (\ref{Rconsts}).
The diagrams for these contributions are shown Fig. \ref{Kinetic}, 
\ref{Darwin},and \ref{SEAgull}, respectively, and the formulas are given in
Appendix A.
For instance, the contribution  associated with the Darwin term is 
given by
\eqn
\Delta {\cal{E}}_d   =  (c_D~c_F-1) \Delta E_d,
\endeqn
since it involves both Darwin and Fermi term.   
The remainder,  the  contributions associated with
the $q^2$ derivative Fermi  and $p'p$ interaction term, requires new
calculation. The diagrams of the $q^2$ derivative Fermi term are analogous  
to the derivative  Fermi term shown in Fig.  \ref{Deri}. 
The diagrams of $p'p$ interaction term is shown in Fig. \ref{E_Rfig}. 
The formulas for both contributions are listed in Appendix B2.
The result of $E_R$ evaluation is summarized  in Table \ref{E_Rtable}.
All terms proportional to $\ln^2(Z\alpha)$ and $\ln(Z\alpha)$ 
can be obtained from $E_B$ and $E_R$. 
We find  
\eqna
(\Delta E)_{\rm log} &=& -{8\over 3} \logsmg 
+ {8\over3}\left(-\ln2+{3\over4}\right)\logmgs  
\nonumber \\
&& - {8\over3}~{11\over 24}\logmg - { 7 \over 2} a_e \logmg~,
\label{E_log}
\endeqna
which agrees with the known result by Layzer \cite{layzer} and 
Zwanziger \cite{zwanziger2}.  Here, the $\ln(m^2/\gamma^2)$ term 
is related to the IR behavior
of the radiative photon and the $\ln(m/\gamma)$ term comes 
>from the threshold singularity due to the exchanged
Coulomb photon. The double logarithm $\ln^2(m/\gamma)$ is 
a consequence of  simultaneous presence of both types of contributions.

The last type of contribution $E_Q$   comes from
the contact term of the NRQED Hamiltonian which is  calculated from 
the QED scattering amplitudes 
exchanging three photons between the electron and the muon, dressed 
by one radiative photon on the electron line. 
>From this one must subtract diagrams which are the KP 
contact term combined with one Coulomb potential 
and the Fermi potential times anomalous magnetic moment combined with 
two Coulomb potentials.  (See Fig. \ref{E_Qfig}.) 
The numerical results evaluated by VEGAS for various photon mass are 
listed in Table \ref{E_Qtable} .
The IR regulator mass  $\lambda$ is kept  finite in the
exchanged photon in the scattering diagram calculation.
The radiative photon is  chosen to be massless. 
This is allowed since the radiative and exchanged photons are separately
gauge invariant. 
The QED scattering contribution depends only on $m/\lambda$ and 
may be written as
\eqn
E_Q = c_1 \logsml + c_2 \logml + c_3 + c_4 {\lambda \over m }+~\cdots~.
\label{EQform}
\endeqn
The coefficients $c_1$, $c_2$, $\cdots$ can be determined by fitting
the formula with the result of numerical integration  \cite{numrecipes}.
Ignoring $c_4$, $c_5$, $\cdots$,  we found 
\eqn
c_1=-2.664~4~(4),~~~c_2=1.171~5~(71),~~~c_3=-8.530~(28). 
\label{c_nfit1}
\endeqn
Actually, the logarithmic part 
\eqn
(E_Q)_{\rm log} = c_1 \logsml + c_2 \logml 
\label{EQform2}
\endeqn
can be determined analytically, noting that the sum of
the contributions $E_B$, $E_S$, $E_R$, and $E_Q$
must be free of both UV  cut-off $ \Lambda$
and IR cut-off $\lambda$.
We find that  
\eqna
 (E_Q)_{\rm log} & = & - (E_S + E_R )_{\rm log}
\nonumber \\
        &=& -{8\over3} \logsml
        -{16\over3}\left( 6\ln2-{9\over 2}\ln3 \right)\logml
        -\left({11\over 9} +{7\over 2}a_e \right) \logml ~.
\label{E_Qexp}
\endeqna
>From this we obtain 
\eqna
          c_1&=& -{8\over3} =-2.666~666~\cdots ~,
\nonumber \\
          c_2&=& 24 \ln3 -32 \ln2 - {11 \over 9} - {7 \over 2 } a_e 
             =1.213~762~9~\cdots~,
\label{c_nexact}
\endeqna
which are in good agreement with the numerical result
(\ref{c_nfit1}).\footnote{If we  choose $c_1=- 8/3$ in Eq. (\ref {EQform}), we obtain
$c_2=1.208~9~(5)$ and $c_3=-8.675~8~(46)$.}
Using the analytic result (\ref{c_nexact}) for $c_1$ and $c_2$, we can
determine the constants $c_3$ and $c_4$ more precisely from 
the numerical result. 
We find \cite{numrecipes}
\eqn
c_3= -8.724~3 \pm 0.001~1,~~~~ c_4=8.688 \pm 0.825~~.
\label{c_3}
\endeqn

Using the constant $c_3 $ in Eq. (\ref{c_3}) together with other 
contributions summarized  in Tables I, II, and III, 
we arrive at the final result for the radiative photon contribution 
to the BES term:
\eqn
\Delta E_{\rm BESph}= 17.122~7~(11)~.
\label{myBESagain}
\endeqn

\section {Discussion}

The value (6) of the $\alpha (Z\alpha)^2$ correction obtained by
Pachucki [13] is in good agreement with our result (4).
However, it is not fully satisfactory in the sense
that details of his calculation seem to disagree with ours
beyond the uncertainty of numerical evaluation.
The numerical value of his low energy contribution $E_L$
(defined by Eq. (16) of [13]) is smaller than our $E_B$ by 0.025 7 (2).
Unfortunately, his mid energy contribution $E_M$ or high energy
contribution $E_H$ cannot be compared directly
with our $E_S$, $E_R$, or $E_Q$.
We find, however, that
the sum $E_M + E_H$
is larger than our sum $E_S + E_R + E_Q$
by 0.024 6(14).
Note that these discrepancies nearly cancel out,
leading to a good agreement of (4) and (6).

In order to compare these results in detail, let us examine $E_L$ more closely.
$E_L$ is the low energy contribution characterized by the parameter $\epsilon$
which is the (UV) cut-off of the frequency $\omega$ of the virtual photon.
In Ref. [13], terms of $E_L$ divergent for $\epsilon \rightarrow \infty$
are separated out and evaluated analytically.
The remainder is evaluated partly analytically and partly numerically.
Adding up the results (24)-(27) of Ref. [13] gives the sum of these terms 
>from which the contribution of
mass renormalization term is not subtracted.
Meanwhile, the mass renormalization effect is subtracted
out in Eqs. (B19), (B20), and (B21).

One way to compare these two methods is to drop the mass
renormalization terms from (B19), (B20), and (B21).
Then evaluate them, for instance, for an NRQED UV cut-off $\Lambda$ satisfying  
$\Lambda m/\gamma^2 = 1$.
Correspondingly,
we choose $\epsilon m/\gamma^2 = 1$
in Pachucki's formula for $E_L$.
We have evaluated both of them numerically.
The agreement between the two integrated results is excellent:
NRQED gives $0.455~763~(0.000~033)$, while Pachucki's formula gives
$0.455~769~(0.000~010)$.

%

In order to identify the leading contributions to the $\alpha(Z\alpha)^2$ correction
we have also performed
an asymptotic expansion of our formulas (without mass renormalization terms)
 in $\gamma^2 / (\Lambda m) $.  The corresponding procedure for Pachucki's $E_L$
is to take the limit of small
%
%
$\gamma^2$ first and
then go to vanishing $\epsilon$.
Both procedures are found to give the
same result for the logarithmically divergent and finite constant terms.
This is as expected since
the mass subtraction term in NRQED has a linear divergence proportional
to $\Lambda m/ \gamma^2 $ without any constant term.
\footnote{In fact,  in the NRQED approach there is also a contribution
proportional
to $\sqrt{ \Lambda m / \gamma^2 }$ from the diagram
$B_{00}$ of Fig.  7.   This term is cancelled out by the same but negative
contribution from the diagram $S_0$ of Fig. 8. }
Note that the corresponding term in Pachucki's method is 
proportional to  $\epsilon m / \gamma^2$.
Thus the  resultant logarithmic  and constant terms
must be of the same form in both $E_B$ and $E_L$, if we identify
$\Lambda = \epsilon $.
%
%
These checks have convinced us that $E_L$ and $E_B$ are
exactly equivalent as far as their analytic properties are concerned.

As the second check, we have also evaluated 
Pachucki's formula for $E_L$, after carrying out an explicit
mass renormalization, by numerical integration for values of $\Lambda (= \epsilon )$
ranging from $10^{2.5} \gamma^2/m$ to $10^6 \gamma^2/m$ and extrapolating the
result to $\Lambda = \infty$.
This method is adopted because precision of numerical integration
deteriorates steadily with increasing $\Lambda$ and direct evaluation of
our formula at $\Lambda = \infty$ becomes impossible.
In spite of our considerable computational effort over several months, however,
the resulting $E_L - E_B$ obtained thus far is not capable of distinguishing
between the values 0 and the difference 0.025 7 mentioned at the beginning
of this section.

In conclusion, we have seen nothing wrong
thus far with the analytic form of either $E_L$ or $E_B$ 
but have been unable to resolve
the apparent numerical discrepancy.
Neither can we rule out an intriguing possibility that Pachucki's 
separation into
low and higher energy parts
is slightly different from ours but the sum is not affected
by how the separation is made.

We publish our result in the present form to avoid undue delay
and to encourage further investigation by others.

The result obtained by Sapirstein for the radiative photon contribution 
was \cite{sapirstein}
\eqn
\Delta E_{\rm BESph}= 15.10~(29).
\label{oldBES}
\endeqn
Concerning the apparent discrepancy between Eqs. (\ref{myBES}) 
and (\ref{oldBES}), Karshenboim pointed out that the latter  
might include the known $\alpha(Z\alpha)^3 \ln(Z\alpha)^{-1}$ correction 
\cite{GPL,karsh2}
\eqna
E_F \alpha (Z\alpha)^3  \logmg 
\left ( 5\ln 2-{ 191 \over 16 }  \right)
&=&E_F \alpha (Z\alpha)^3  \logmg 
 ( -8.471~764 \cdots)
\nonumber \\
&=& E_F {\alpha(Z\alpha)^2 \over \pi }(-0.955 \cdots) 
\label{highestimate}
\endeqna
in the relativistic bound state formalism adapted
in Ref. \cite{sapirstein}.
%
%
%
%
%
%
We wrote the second line of (\ref{highestimate}) to emphasize that, 
numerically speaking, this
contribution is of the same order of magnitude as the $\alpha(Z\alpha)^2$
term.

Eq. (\ref{oldBES}) will contain in addition the non-logarithmic part of the
$\alpha(Z\alpha)^3$ correction. If terms of order $\alpha(Z\alpha)^4$ or higher 
can be ignored, this contribution 
may be estimated from Eqs. (\ref{myBES}), (\ref{oldBES}),
and (\ref{highestimate}) to be
\eqn
E_F \alpha (Z\alpha)^3 ( -37 \pm 13 ). 
\label{highestimate2}
\endeqn
In order to see whether this is plausible or not,
let us note, based on the NRQED analysis,  that 
the  non-logarithmic term of the $\alpha(Z\alpha)^3$  term 
may be smaller by only a factor of $\logmg\sim 5$ compared  with the
leading term  Eq.(\ref{highestimate}):
\eqn
E_F \alpha (Z\alpha)^3  
\left ( 5\ln 2-{ 191 \over 16 }  \right) \times  {\cal{O}}(1)
=E_F \alpha (Z\alpha)^3   
 ( -8.471~764 \cdots) \times {\cal{O}}(1).
\label{highestimate3}
\endeqn
Thus the estimate (\ref{highestimate2}) is somewhat large but  not
inconsistent with our NRQED estimate.

The correction due to one radiative photon
to the hyperfine structure  was computed recently
for various atomic number $Z$ \cite{persson,BCS}
without expanding in powers of $Z\alpha$.\footnote{We thank P. Mohr 
and B. Taylor for drawing our attention to 
Ref. \cite{persson} and discussing their results. }
Thus their results for $Z=1$ must contain corrections of order 
$\alpha(Z\alpha)^n$ with  $n=0, \cdots ,\infty $. 
Unfortunately, the results at $Z=1$ do not have enough accuracy 
in order to extract the non-logarithmic contribution  of
the $\alpha(Z\alpha)^2$ correction, namely the BES term. 
Their results for high $Z$  are not sufficiently accurate 
either for a reliable extrapolation  to $Z=1$.

In this circumstance it seems that  
only  a direct evaluation of the pure $\alpha(Z\alpha)^3$ 
term will clarify the  ambiguity in the theoretical calculation 
concerning this term.
This problem will be treated in our subsequent paper \cite{NK3}.

To summarize, we have calculated the $\alpha(Z\alpha)^2$ correction to the
hyperfine splitting to the ground state muonium based on the NRQED.
Including the logarithmic terms, the total contribution of this order 
becomes
\eqna
\Delta \nu[\alpha(Z\alpha)^2]
& = &{\alpha(Z\alpha)^2 \over \pi } E_F  \biggl [
                {8 \over 15} \logmg -{8\over15}\ln2
                +{214\over 225}-{4\over 5}  
\nonumber \\
&& - {8\over 3} \logsmg
+ {8\over3}\left(-\ln2+{3\over4}\right)\logmgs
\nonumber \\
&& -\left ({8\over3}~{11\over 24} + { 7 \over 2} a_e \right)\logmg~
              +17.122~7~(11)   ~\biggr ]
\nonumber \\
&=&
 E_F {\alpha(Z\alpha)^2 \over \pi}
                \biggl [ -{8 \over 3} \logsmg
\nonumber \\
&&  + \left( - {8\over3}   \ln4 + {37\over36} + {8\over 15}
                       \right ) \logmg
              +  16.904~2~(11)~\biggr ]~~~,
\endeqna
where $a_e=1/2$.
Note that the  uncertainty due to the numerical integration is now
%
%
%
%
%
%
reduced drastically to match the expected accuracy of the
forthcoming experiment.

Now that we have completed the evaluation of the $\alpha^n (Z\alpha)^{3-n}$, 
n=1,2,3, terms, 
it seems about time to discuss the numerical value of the theoretical 
prediction of the hyperfine splitting of the muonium. 
However,  the leading logarithmic corrections of  order
$\alpha^n(Z\alpha)^{4-n},n=0,1,2,3,~$  turn out to be numerically 
of the same order of magnitude as some $\alpha^3$ corrections. 
Thus, we postpone comparison with experiment  to the next paper in which
we will treat these higher order logarithmic corrections.

\acknowledgments

We thank G. P. Lepage, P. Labelle, late D. R. Yennie, 
J. Sapirstein, P. Mohr, B. Taylor, and K. Pachucki  for useful discussions.
This research is supported in part by the U. S. National Science
Foundation.
M. N.  thanks  Department of Physics and Astronomy and
Center for Computational Sciences
at the University of Kentucky for their support.
Part of numerical work was conducted at the Cornell National
Supercomputing Facility, which receives major funding from the US
National Science Foundation and the IBM Corporation, with additional
support from New York State and members of the Corporate Research
Institute.
Some of the numerical work was  carried out
at the Center for Computational Sciences of  the University of Kentucky 
and at the Computer Center of  Nara Women's University.   
This work is also  supported by National Laboratory for High Energy Physics, 
Japan,
as a KEK Supercomputer project (Project Number 11).

\appendix

\def\fermion#1#2{ {-2m \over \vec{#1}\,^2 + {#2}^2 }}
\def\fermis#1{ {-2m \over \vec{#1}\,^2  }}
\def\fermip#1{ {\vec{#1}\,^2 \over \vec{#1}\,^2 +\lambda^2 }}
\def\coulomb#1#2{ {-4\pi Z\alpha \over |\vec{#1}-\vec{#2}|^2  }}
\def\coulombs#1#2{ {-4\pi Z\alpha \over |\vec{#1}-\vec{#2}|^2 +\lambda^2}}
\def\combs#1{ {-4\pi Z\alpha \over \vec{#1}\,^2 +\lambda^2}}
\def\wavef{ { (8\sqrt{\pi \gamma^5})^2 \over  
(\vec{p}\,^2+\gamma^2)^2 (\vec{r}\,^2+\gamma^2)^2} }
\def\wavefc#1#2{ { (8\sqrt{\pi \gamma^5})^2 \over  
(\vec{#1}\,^2+\gamma^2)^2 (\vec{#2}\,^2+\gamma^2)^2} }
\def\pqlrint{\int\! { d^3p d^3q d^3 l d^3 r \over (2\pi)^{12} } }
\def\pqrint{\int\! { d^3p d^3q  d^3 r \over (2\pi)^9} }
\def\print{\int\! { d^3p  d^3 r \over (2\pi)^6} }
\def\pqint{\int\! { d^3p  d^3 q \over (2\pi)^6} }
\def\pint{\int\! { d^3p   \over (2\pi)^3} }
\def\kint{\int^{\Lambda}_0 \! dk }
\def\kintv{\int^{\Lambda}_0 \! dk k}
\def\overall{ 2E_F{\pi \over \gamma^3}~{2\alpha \over 3 \pi }}
\def\overalls{ E_F~{2\alpha \over 3 \pi }}
\def\pola#1#2{ {\vec{#1}\cdot\vec{#2} \over m^2 }}
\def\DE#1#2{ \Delta E[\rm{Fig.}\ref{#1}({#2})]}

\section{Calculation of the  $(Z\alpha)^2$ Correction}

\subsection{List of contact terms}

In addition to the contact term (\ref{V_1c-k}) against the relativistic kinetic energy term, there are also
contact terms  against the Darwin term, the derivative Fermi
term and the seagull term determined from the two photon exchange scattering
amplitudes. They are given by  
\eqna
V_{1c-d}  & = &-2 \int^{\Lambda}\!\frac{d^{3}k}{(2\pi)^{3}}
              V_F(0,\vec{k}) \fermis{k} V_D(\vec{k},0)~,
\\
V_{1c-w} & =& -2  \int^{\Lambda}\!\frac{d^{3}k}{(2\pi)^{3}}
              V_W(0,\vec{k}) \fermis{k}
              V_C(\vec{k},0)~,
\\
V_{1c-s}  &=& -2  \int^{\Lambda}\!\frac{d^{3}k}{(2\pi)^{3}}
                 V_S(0,0,\vec{k}),
\label{V_1others}
\endeqna
where the potentials $V_F, V_D, V_W, V_C,$ and $V_S$ are defined 
by (\ref{V_F}), 
(\ref{V_D}), (\ref{V_W}), (\ref{V_C}), and (\ref{V_S}), respectively.

>From the three photon exchange scattering amplitude, we find, in addition to
(\ref{V_2c-k(1)}),  the
following contact terms against the relativistic kinetic term:
\eqna
V_{2c-k(2)} & = &-2 \int^{\Lambda}\! \frac{d^{3}k}{(2\pi)^{3}}
                      \int\! \frac{d^{3}l}{(2\pi)^{3}}
                  V_F(0,\vec{k}) \fermis{k}
                  V_K(\vec{k}) \fermis{k}
                  V_C(\vec{k},\vec{l}) \fermis{l}
                  V_C(\vec{l},\vec{0})~, 
\label{V_2c-k(2)}
\\
V_{2c-k(3)} & = &-2 \int^{\Lambda}\! \frac{d^{3}k}{(2\pi)^{3}}
                      \int\! \frac{d^{3}l}{(2\pi)^{3}}
                  V_C(0,\vec{k}-\vec{l})  {-2m \over |\vec{k}-\vec{l}|^2} 
                  V_K(\vec{k}-\vec{l}) 
                  {-2m \over |\vec{k}-\vec{l}|^2} 
\nonumber \\       
&& ~~~~~~~~~~~~~~~~~~\times 
                  V_F(\vec{k}-\vec{l},\vec{l}) \fermis{l}
                  V_C(\vec{l},\vec{0}) ~. 
\label{V_2c-k(3)}
\endeqna

\subsection{ The Relativistic Kinetic Energy Term }

\begin{figure}  
\centerline{\epsfbox{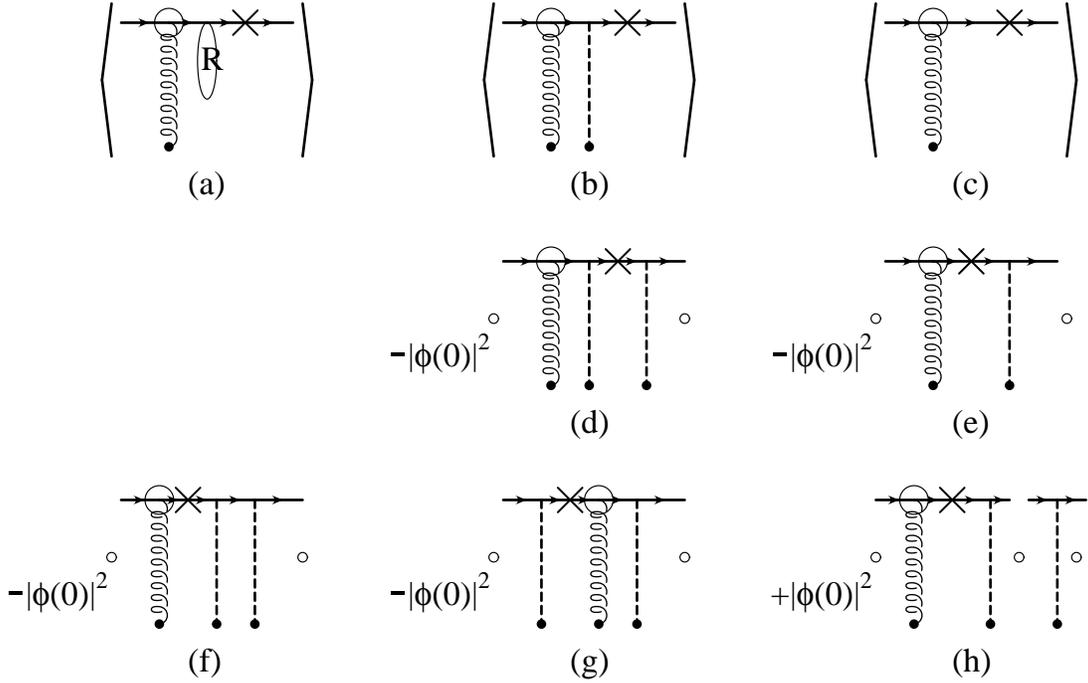}}
\vspace{2ex}
\caption{Relativistic kinetic term diagrams.
The brackets $\langle$~~$\rangle$ indicate that the diagram is
evaluated with the bound state wave function.
The contact term diagram 
is shown in the right below  of the corresponding bound state.
The time-reversed diagrams are not shown.
R in (a) represents two or more photon exchange part of the
Coulomb Green's function.
\label{Kinetic}}
\end{figure}

\begin{figure} 
\centerline{\epsfbox{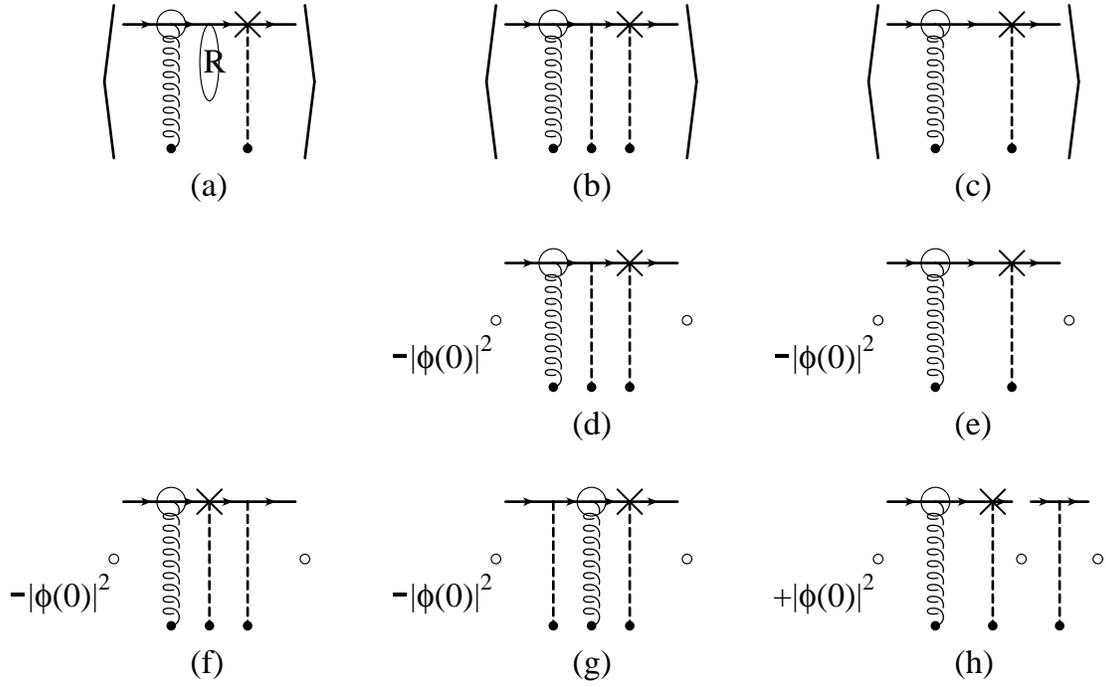}}
\vspace{2ex}
\caption{Darwin term diagrams. \label{Darwin}}
\end{figure}

\begin{figure} 
\centerline{\epsfbox{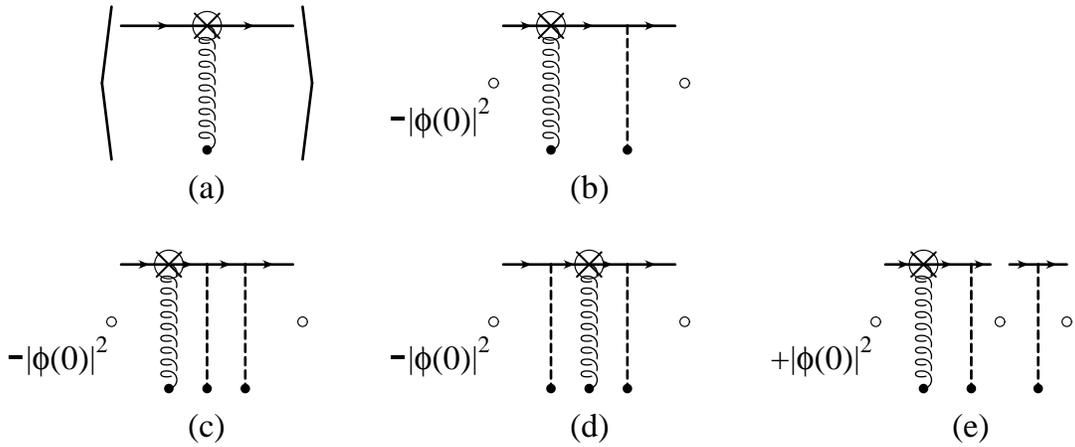}}
\vspace{2ex}
\caption{Derivative Fermi term diagrams. \label{Deri}}
\end{figure}

\begin{figure} 
\centerline{\epsfbox{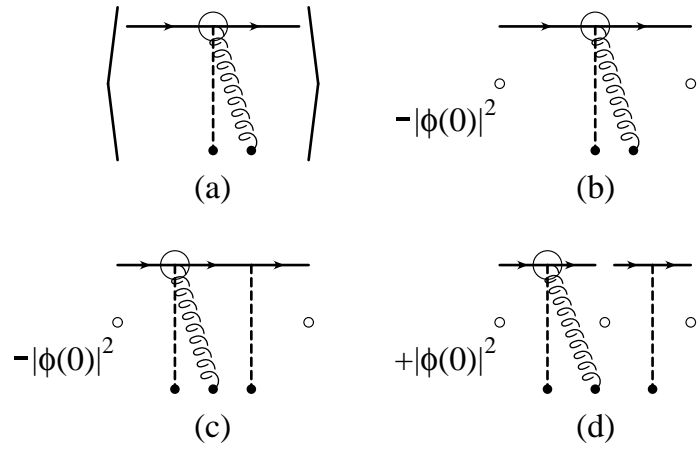}}
\vspace{2ex}
\caption{Seagull term diagrams. \label{SEAgull}}
\end{figure}

\begin{figure}
\centerline{\epsfbox{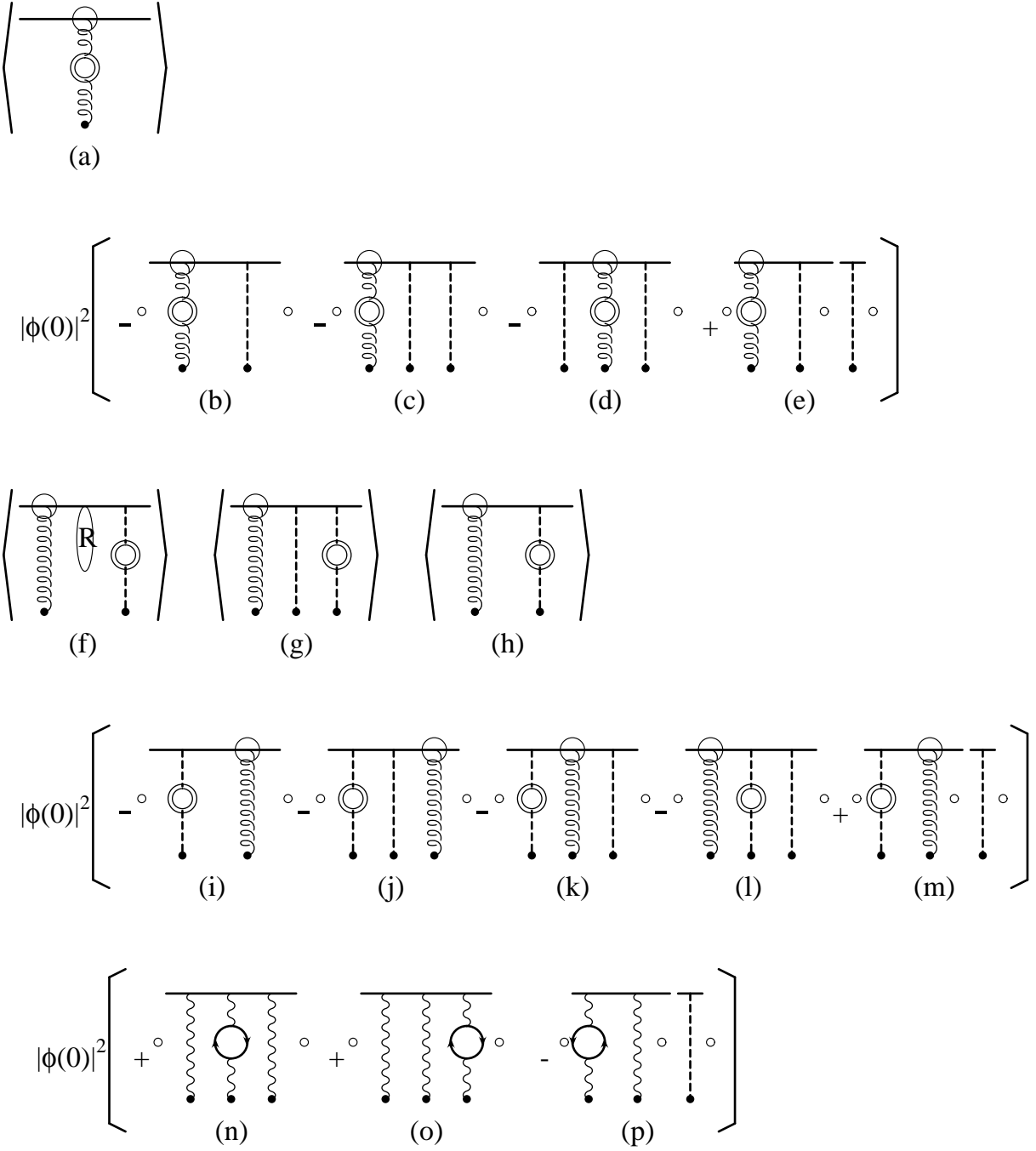}}
\caption{ Vacuum polarization diagrams contributing to 
the $\alpha(Z\alpha)^2$ correction.
\label{BSvp}}
\vspace{3ex}
\end{figure}

The correction coming from the relativistic kinetic energy term
shown in Fig. \ref{Kinetic} 
is calculated in second-order perturbation theory given 
in (\ref{perturb}). 

The diagram of Fig. \ref{Kinetic}(a) 
with two or more photons exchanged leads to  a finite result
\eqna
\DE{Kinetic}{a} & = & 2 
             \pqlrint
             \wavef
\nonumber  \\
        & \times &   \langle V_F(\vec{r},\vec{l}) \rangle 
       \biggl (\frac{-64\pi}{Z\alpha\gamma^{4}} \biggr )
      \tilde{R}' (\vec{l},\vec{q};E_0)
       V_K(\vec{q},\vec{p})
\nonumber  \\
         & = & E_{F}(Z\alpha)^{2} \frac{(-8)}{\pi}
               \int^{\infty}_{0}
                \frac{\gamma q^{6}dq}{(q^{2}+\gamma^{2})^{4}}
           \biggl [\ln 2 - \frac{5}{2} + \frac{\gamma}{q} 
              \tan^{-1}\frac{q}{\gamma}
\nonumber \\
&& -\frac{1}{2}\ln (1+\frac{q^{2}}{\gamma^{2}}) 
             + \frac{4\gamma^{2}}{q^{2}+\gamma^{2}}  \biggr ]
\nonumber  \\
         & = & E_{F}(Z\alpha)^{2} \frac{51}{16}~,
\label{E_kR}
\endeqna
where the factor 2 accounts for two contributing diagrams.
The photon mass $\lambda$ is set to zero  because
the bound state is slightly off-shell and  free from IR
divergence. 
The brackets $\langle \cdots  \rangle$ mean that we take the difference
between the expectation values with respect to the 
spin $J=1$ state and the spin $J=0 $ state. 
$\tilde{R}'$ is the two or more photon exchange part $\tilde{R}$
of Green's function $\tilde{G}_0$ given by Eq. (\ref{green1}) minus 
the  pole contribution of the ground state:
\eqn
\tilde{R}' \equiv 
\lim_{E \rightarrow E_0} \left( \tilde{R}-
\frac{|\psi_0 \rangle  \langle\psi_0 |} {E-E_0 } \right)~.
\endeqn
The explicit form of $\tilde{R}'$ can be found in Eq. (28) of Ref.\cite{KN1}. 
Integration over $\vec{l}$ and $\vec{q}$ in  (\ref{E_kR})
can be carried out using
%
%
%
%
%
%
the identity for $\tilde{R}'$ given in Ref. \cite{CLold}: 
\eqn
       \int\!{d^3l\over(2\pi)^3} 
       \int\!{d^3q\over(2\pi)^3} 
         \tilde{R}'(\vec{l},\vec{q};E_0) f(|\vec{q}|)=
              -{\gamma^7\over 16\pi^3}\int^{\infty}_{0}
              dq{q^2\over(q^2+\gamma^2)^2} R(q)f(q)~,
\endeqn
where 
\eqn
R(q)=\ln 2 - {5\over2} + {\gamma \over q}\tan^{-1}
                     \biggl ({q\over\gamma}\biggr)
    -{1 \over 2}\ln \biggl (1+{q^2\over\gamma^2} \biggr) 
    + {4\gamma^2\over q^2+\gamma^2}~.
\endeqn

Unlike the two or more photon exchange part,
the zero and  one photon exchange parts of the  Green function
give rise to linear and logarithmic divergences, respectively.  
These divergences must be taken care of by the NRQED contact terms.
Thus it is convenient to treat the calculation of the  bound state  
and the corresponding contact terms together, since they 
have the same UV behavior.

For the one photon exchange part  
the bound state calculation gives
\eqna
\DE{Kinetic}{b} & = &  2 \pqlrint \wavef
          \langle V_F(\vec{r},\vec{l}) \rangle  \fermion{l}{\gamma}
\nonumber \\
      && \times    V_C(\vec{l},\vec{q})   \fermion{q}{\gamma}
           V_K(\vec{q},\vec{p})   
\nonumber \\
         & = &  E_{F}(Z \alpha)^{2} \left [- 2\ln 2 
               +2 \logLg 
              -\frac {13}{16} \right ]~,                   
\endeqna
where $\Lambda$ is the UV cut-off.  It is put to  infinity 
for the finite parts.

The corresponding contact term should have the same structure and
we find it among the two-loop contact terms. It is 
one Coulomb photon exchange between the Fermi and 
the relativistic kinetic energy potentials.
(See Fig. \ref{Kinetic} (d).)
Since this  is a contact potential,  the first order perturbation 
theory with the potential becomes the wave function at the origin
squared $ |\phi(0)|^2=\gamma^3/\pi $  times the minus sign of
its scattering amplitude.
To avoid the IR singularity caused by the on-shell condition, 
we put the infinitesimal photon mass $\lambda$
in the photon propagators.
This leads to
\eqn
\DE{Kinetic}{d}  =   |\phi(0)|^2  \langle  V_{2c-k(1)} \rangle   
           =  E_{F}(Z\alpha)^{2} \left [ -2\logLl + 2\ln3  \right ]~,
\endeqn
where $V_{2c-k(1)}$ is given by (\ref{V_2c-k(1)}).
This  $\ln\Lambda$ cancels that of  $\DE{Kinetic}{b}$.

The zero-photon exchange part of the bound state theory 
can be calculated in a similar way. (See Fig.\ref{Kinetic}(c).)
One subtle thing takes place in the calculation of the zero photon
exchange part, because  linear divergence causes  the integrated 
result to depend on how it is  parameterized. 
If the regulated momentum is shifted, it may  give an 
additional finite piece to the answer. 
In order to get rid of the uncertainty due to linear
divergence, we will consider the bound state calculation
and the contact term calculation together and put the cut-off 
$\Lambda$  in the corresponding transverse photon line \cite{pat}. 
We choose  the  momentum $\vec{k}$  of the transverse
exchanged photon line 
to have the  cut-off $\Lambda$ 
throughout the  whole calculation, since it is easily recognized.
Keeping the cut-off $\Lambda$ at finite value, we  get 
\eqna
\DE{Kinetic}{c} & = & 
           2 \int\!{d^{3}\vec{p}\over(2\pi)^{3}}
             \int^{\Lambda}{d^{3}\vec{k}\over(2\pi)^{3}}
              \frac{(8\sqrt{\pi\gamma^{5}})^{2}}
              {(|\vec{p}+\vec{k}|^{2}+\gamma^{2})^{2}
              (\vec{p}\,^2+\gamma^{2})^{2}}  
\nonumber \\
           &\times &  V_F(\vec{p}+\vec{k},\vec{p}) \fermion{p}{\gamma}
             V_K(\vec{p})
\nonumber \\
        & = & E_{F}(Z\alpha)^{2} \left [ 2\frac{\Lambda}
           {\gamma\pi}-\frac{23}{8} \right ]~.
\endeqna
The corresponding contact term gives 
\eqn
\DE{Kinetic}{e}  =  \frac{\gamma^{3}}{\pi} 
             \langle V_{1c-k} \rangle  
  =  E_{F}(Z\alpha)^{2}\left [-2 \frac{\Lambda}{\gamma\pi} \right ]~,
\endeqn
where $V_{1c-k}$ is given by (\ref{V_1c-k}).
In these calculations,  we have used the expansion 
\eqn
 \tan^{-1}\biggl(\frac{\Lambda}{\lambda}\biggr)=
                       \frac{\pi}{2} - \frac{\lambda}{\Lambda}
   + {1\over 3}\biggl(\frac{\lambda}{\Lambda}\biggr)^3
   +O\biggl(\biggl(\frac{\lambda}{\Lambda}\biggr)^{5}\biggr)~.
\endeqn
Clearly, the one-loop contact term kills the linear $\Lambda$ divergence 
in $\DE{Kinetic}{c}$.

The other  two-loop contact terms are linearly divergent
due to the threshold singularity.
The diagram Fig. \ref{Kinetic}(f)  gives
\eqn
\DE{Kinetic}{f}  = \frac{\gamma^{3}}{\pi}\langle V_{2c-k(2)} \rangle 
        =  E_{F}(Z\alpha)^{2}  \left [- 4\frac{\Lambda}{\lambda\pi}-
          2 \ln\frac{2}{3} + 3  \right ]~,
\endeqn
where $V_{2c-k(2)}$ is given by (\ref{V_2c-k(2)}).
The diagram Fig.\ref{Kinetic}(g)  has an expression different from 
the diagram Fig. \ref{Kinetic}(f), but their integrals are identical:
\eqn
\DE{Kinetic}{g}  = \frac{\gamma^{3}}{\pi} \langle V_{2c-k(3)} \rangle 
 =  E_{F}(Z\alpha)^{2}  \left [ -4\frac{\Lambda}{\lambda\pi}-
          2 \ln\frac{2}{3} + 3  \right ]~,
\endeqn
where $V_{2c-k(3)}$ is given by (\ref{V_2c-k(3)}).

The remainder is the contact term that consists of   a one-loop
contact term and one Coulomb photon exchange:
\eqn
\DE{Kinetic}{h}  = \frac{\gamma^{3}}{\pi} \langle V_{2c-k,1loop^2} \rangle 
 =  E_{F}(Z\alpha)^{2}   \left [ 8\frac{\Lambda}{\lambda\pi} -6  \right ]~,
\endeqn
where $V_{2c-k,1loop^2}$ is given by (\ref{V_2c-kloop}).

Summing up all the contributions relating to the kinetic term correction, 
we get 
\eqn
\Delta E_{k} =  E_{F}(Z\alpha)^{2}  \left [  -\frac{1}{2}-6\ln\frac{2}{3}
    +2\loglg  \right ]~.      \label{Ek}
\endeqn
Each linearly divergent diagram could have a value which 
depends on the regularization method,
but the  sum remains the same because of  gauge invariance.

\subsection{ The Darwin Term}

The computation method of the Darwin term 
shown in Fig. \ref{Darwin} is almost 
identical with  that of the kinetic term. 
The bound state calculation is  also from the second order
perturbation theory.
The number of graphs has been already taken into account in the 
following equations:
\eqna
\DE{Darwin}{a} & = &2 \pqlrint \wavef
\nonumber \\
  &\times &     \langle  V_F(\vec{r}, \vec{l}) \rangle
\left(\frac{-64\pi}{Z\alpha\gamma^{4}}\right )\tilde{R}'(\vec{l},\vec{q};E_0)
           V_D(\vec{q},\vec{p} )
\nonumber \\
       & = & E_{F}(Z\alpha)^{2} \left [- \frac{3}{2} \right ]~,
\label{Eda}
\endeqna

Similarly, we find
\eqna
\DE{Darwin}{b} & = & E_{F}(Z\alpha)^{2} \left [-\logLg +\ln2 \right ]~,
\\
 \DE{Darwin}{c} & = &E_{F}(Z\alpha)^{2} 
             \left [ -\frac{\Lambda}{\gamma \pi}+1 \right ]~.
\endeqna

The contact terms give the following contributions:
\eqna
\DE{Darwin}{d} & = &-2 \frac{\gamma^{3}}{\pi}\int^{\Lambda}
              \frac{d^{3}k}{(2\pi)^{3}}  
              \langle    V_F(0,\vec{k}) \rangle 
             \frac{-2m}{\vec{k}\,^2} V_D(\vec{k},0)
\nonumber \\
           & = &E_F(Z\alpha)^{2}\frac{\Lambda}{\gamma\pi} ~,
\\
\DE{Darwin}{e} & = & 
           E_{F}(Z\alpha)^{2} \left [ \logLl-\ln3  \right ]~,
\\
\DE{Darwin}{f} & = & 
           E_{F}(Z\alpha)^{2} \left [2 \frac{\Lambda}{\lambda\pi}
             +\ln\frac{2}{3} -1  \right ]~,
\\
\DE{Darwin}{g} & = & 
           E_{F}(Z\alpha)^{2} \left [2 \frac{\Lambda}{\lambda\pi}
             +\ln\frac{2}{3} -\frac{3}{2}  \right ]~,
\\
\DE{Darwin}{h} & = & 
     E_{F}(Z\alpha)^{2} \left [-4\frac{\Lambda}{\lambda\pi}+3 \right ]~.
\label{Edz}
\endeqna

The sum of the contributions (\ref{Eda}) -- (\ref{Edz}) is
\eqn
\Delta E_d=  E_{F}(Z\alpha)^{2}  \left [
 -\ln \left (\frac{\lambda}{\gamma} \right) 
+ 3\ln \frac{2}{3}   \right ]~.    \label{Ed}
\endeqn

\subsection{ The Derivative Fermi Term}

The corrections involving the derivative Fermi term 
shown in Fig. \ref{Deri} 
are calculated in the first order perturbation theory.
We obtain
\eqna
\DE{Deri}{a} & = &   \int\frac{d^{3}p}{(2\pi)^3}
                       \int^{\Lambda}\frac{d^{3}k}{(2\pi)^3}
              \frac{(8\sqrt{\pi\gamma^{5}})^{2}}
              {(|\vec{p}-\vec{k}|^{2}+\gamma^{2})^{2}
               (\vec{k}\,^2+\gamma^{2})^{2}}  
          \langle V_W(\vec{p},\vec{p}-\vec{k})  \rangle 
\nonumber \\
   & = & E_{F}(Z\alpha)^{2} 
       \left [-2\frac{\Lambda}{\gamma\pi}+\frac{5}{2} \right ] ~,
\label{Ewa}  
\\
\DE{Deri}{b} & = &
          E_{F}(Z\alpha)^{2} 2\frac{\Lambda}{\gamma \pi} ~,
\\
\DE{Deri}{c} & = &
          E_{F}(Z\alpha)^{2}
          \left [4\frac{\Lambda}{\lambda \pi} +2\ln\frac{2}{3} -3 \right ] ~,
\\
\DE{Deri}{d} & = & 
          E_{F}(Z\alpha)^{2}
  \left [4\frac{\Lambda}{\lambda \pi} +2\ln\frac{2}{3} -3 \right ] ~,
\\
\DE{Deri}{e} & = & 
   E_{F}(Z\alpha)^2   \left [-8 {\Lambda \over \lambda\pi} + 6   \right ]~.
\label{Ewz}
\endeqna

The sum of the contributions (\ref{Ewa}) --(\ref{Ewz}) is
\eqn
\Delta E_w=  E_{F}(Z\alpha)^{2}   
 \left [ 4\ln \frac{2}{3}  +\frac{5}{2}  \right ]~.    \label{Ew}
\endeqn

\subsection{ The Seagull Term}

The diagrams involving the seagull term are shown in Fig. \ref{SEAgull}. 
The bound state calculation is also carried out in the first order 
perturbation theory.
The results are
\eqna
\DE{SEAgull}{a} & = & 
             2 \int\frac{d^{3}p}{(2\pi)^3}
              \int^{\Lambda}\frac{d^{3}k}{(2\pi)^3}
              \int \frac{d^{3}l}{(2\pi)^3}
              \frac{(8\sqrt{\pi\gamma^{5}})^{2}}
              {(|\vec{p}-\vec{k}|^{2}+\gamma^{2})^{2}
              (|\vec{p}-\vec{l}|^{2}+\gamma^{2})^{2} }
\nonumber \\ 
& \times &
          \langle V_S( \vec{p}-\vec{l},\vec{p}-\vec{k},\vec{l}) \rangle
\nonumber \\
         & = & E_{F}(Z\alpha)^{2} \left [\frac{\Lambda}{\gamma\pi}
                                -\logLg+\ln2 \right ]~,
\label{Esa}
\\
\DE{SEAgull}{b} & = & 
          E_{F}(Z\alpha)^{2} \left[ - \frac{\Lambda}{\gamma\pi} \right ]~,
\\
\DE{SEAgull}{c} & = &
         E_{F}(Z\alpha)^{2} \left [-4\frac{\Lambda}{\lambda\pi}
             +\logLl
             +{5\over2}+\ln3 -2\ln2  \right ]~,
\\
\DE{SEAgull}{d} & = & 
 E_{F}(Z\alpha)^{2}   \left [4\frac{\Lambda}{\lambda\pi}-3  \right ]~.
\label{Esz}
\endeqna

The sum of the contributions (\ref{Esa}) -- (\ref{Esz}) is
\eqn
\Delta E_s=  E_{F}(Z\alpha)^{2}  \left [
 -\loglg  - \ln \frac{2}{3}  -\frac{1}{2}  \right ]~.   \label{Es}
\endeqn

Adding up Eqs. (\ref{Ek}), (\ref{Ed}), (\ref{Ew}), and (\ref{Es}) we obtain 
the Breit $(Z\alpha)^2$ term 
\eqn
\Delta E_{\rm Breit} = {3 \over 2} E_{F}(Z\alpha)^{2}. 
\endeqn

\section{Calculation of the  $\alpha(Z\alpha)^2$ Correction}

\subsection{Vacuum Polarization Contribution}

First let us calculate  the NRQED corrections related to 
the vacuum polarization insertion in the transverse exchanged photon. 
The bound state calculation is the first order perturbation theory with
the NRQED potential $V_{\rm tvp}$ of (\ref{Vtvp}):
\eqna
\DE{BSvp}{a} & = &E_F {\pi \over \gamma^3  } { \alpha \over 15 \pi } 
\int {d^3p \over (2\pi)^3 }\int^{\Lambda} {d^3k \over (2\pi)^3}
{(8\sqrt{\pi \gamma^5})^2 \over (\vec{p}\,^2 + \gamma^2)^2
(|\vec{p}+\vec{k}|^2 + \gamma^2)^2 }{\vec{k}\,^2 \over m^2 }
\nonumber \\
& =& E_F {\alpha(Z\alpha)^2 \over \pi} 
{8\over 15} \left [{\Lambda \over \gamma \pi } -{3 \over 2}  \right ] ~.
\endeqna
Similarly, for the scattering  diagrams of Fig.\ref{BSvp}(b) -- (e), 
we obtain 
\eqna
\DE{BSvp}{b}
& = &  E_F { \alpha (Z\alpha)^2 \over  \pi } {-8 \over 15}
{\Lambda \over \gamma \pi }~,
\\
\DE{BSvp}{c}
& = & E_F { \alpha (Z\alpha)^2 \over  \pi } {8 \over 15}
\left [ -2{\Lambda \over \lambda \pi }- 2\ln {2\over3} + {17 \over 12} \right ]~,
\\
\DE{BSvp}{d}
& = &  E_F { \alpha (Z\alpha)^2 \over  \pi } {8 \over 15}
 \left [-2{\Lambda \over \lambda \pi}-3\ln 2+{3\over2}\ln 3+{17 \over 6} 
\right ]~,
\\
\DE{BSvp}{e}
& = &  E_F { \alpha (Z\alpha)^2 \over  \pi } {8 \over 15}
 \left [4 {\Lambda \over \lambda \pi}-{15 \over 4} \right ]~.
\endeqna

Next let us calculate the contribution coming from the NRQED potential
representing the vacuum polarization insertion in the Coulomb photon
given by the potential $V_{\rm cvp}$ of Eq. (\ref{Vcvp}).
The bound state calculation is in the second order perturbation theory.
The two- or more-photon-exchange part of the Coulomb Green's function
gives the correction 
\eqna
\DE{BSvp}{f} &=& 2 E_F {\pi \over \gamma^3  } 
{-4\pi (Z\alpha) \alpha \over 15 m^2 \pi } 
\pqlrint
\wavef
\nonumber \\
& \times &
{ -64 \pi \over (Z\alpha) \gamma^4 } \tilde{R}'(\vec{q},\vec{l};E_0)
\nonumber \\
& =&  E_F {\alpha(Z\alpha)^2 \over \pi}{4 \over 5}~. 
\endeqna
Similarly,
the one-photon-exchange part gives  
\eqna
\DE{BSvp}{g}  
& =&  E_F {\alpha(Z\alpha)^2 \over \pi}{8 \over 15} 
  \left [ \logLl - \ln2  \right ]~.
\endeqna
The zero-photon-exchange part gives  
\eqna
\DE{BSvp}{h}
&=&  E_F {\alpha(Z\alpha)^2 \over \pi}{8 \over 15} 
  \left [ {\Lambda \over \gamma \pi} - 1  \right ]~.
\endeqna
The corresponding scattering  diagrams give the following
corrections:
\eqna
\DE{BSvp}{i}& =& 
-2 E_F 
{-4\pi (Z\alpha) \alpha \over 15 m^2 \pi } 
\int^{\Lambda} {d^3k \over (2\pi)^3}
{ \vec{k}\,^4 \over (\vec{k}\,^2 + \lambda^2)^2 }
\fermis{k}  \fermip{k}
\nonumber \\
&=&
E_F {\alpha(Z\alpha)^2 \over \pi}{-8 \over 15}{\Lambda \over \pi \gamma}~.
\endeqna
Similarly, we obtain
\eqna
\DE{BSvp}{j} &=&
E_F {\alpha(Z\alpha)^2 \over \pi}{-8 \over 15} 
\left [ \logLl-\ln3-{1\over6} \right ]~,
\\
\DE{BSvp}{k}
&=&
E_F {\alpha(Z\alpha)^2 \over \pi}{-8 \over 15}
 \left [2 {\Lambda\over \lambda \pi} +\ln{2\over3} 
-{23\over 12} \right ]~,
\\
\DE{BSvp}{l} &=&
E_F {\alpha(Z\alpha)^2 \over \pi}{-8 \over 15}
 \left [2{\Lambda\over \lambda \pi} +2\ln{2\over3} -{21\over12} \right ]~,
\\
\DE{BSvp}{m}& =& 
E_F {\alpha(Z\alpha)^2 \over \pi}{8 \over 15}
 \left [4 {\Lambda\over \lambda \pi} -{15\over4} \right ]~.
\endeqna
The sum of $\DE{BSvp}{a}$ $\cdots$ $\DE{BSvp}{m}$ gives the NRQED contribution 
of $\Delta E_{\rm NRQED vp}$ of Eq. (\ref{Enrqedvp}).

The remainder is the QED scattering diagrams contribution.
The QED three-photon-exchange skeleton diagram gives the
hyperfine splitting contribution:
\eqn
\Delta E_{skl}=
E_F{(Z\alpha)^2 \over \pi^2}
\int_0^{\infty} dp\int_0^{\infty} dq \int^1_{-1}d(cos\theta)
{ 16 ( \vec{p}\,^2 - \vec{p}\cdot\vec{q} + \vec{q}\,^2 )
\over (\vec{p}\,^2 +\lambda^2)(|\vec{p}-\vec{q}|^2+\lambda^2)
       (\vec{q}\,^2 +\lambda^2)} ~.
\endeqn
Thus the vacuum polarization insertion in the middle exchanged photon
gives 
\eqna
\DE{BSvp}{n} &=&
E_F {(Z\alpha)^2 \over \pi^2 } \int ^1_0 dt \rho_2(t) 
\int_0^{\infty} dp\int_0^{\infty} dq \int^1_{-1}d(cos\theta)
{ |\vec{p}-\vec{q}|^2 \over  |\vec{p}-\vec{q}|^2 +\lambda^2 }
\nonumber \\    
& & \times 
{ 16 ( \vec{p}\,^2 - \vec{p}\cdot\vec{q} + \vec{q}\,^2 )
\over (\vec{p}\,^2 +\lambda^2)(|\vec{p}-\vec{q}|^2+4m^2(1-t^2)^{-1})
       (\vec{q}\,^2 +\lambda^2)} ~,
\endeqna
where $\rho_2(t)$ is the second-order photon spectral function
given by 
\eqn
\rho_2(t)={\alpha \over \pi} { t^2(1-{1\over3}t^2 ) \over 1-t^2 }~.
\endeqn
The vacuum polarization insertion in the outermost exchanged photon leads to
\eqna
\DE{BSvp}{o} & =&
2 E_F{(Z\alpha)^2 \over \pi^2 } \int^1_0 dt \rho_2(t) 
\int_0^{\infty} dp\int_0^{\infty} dq \int^1_{-1}d(cos\theta)
{ \vec{p}\,^2 \over  \vec{p}\,^2 +\lambda^2 }
\nonumber \\ 
&& \times
{ 16 ( \vec{p}\,^2 - \vec{p}\cdot\vec{q} + \vec{q}\,^2 )
\over (\vec{p}\,^2 +4m^2(1-t^2)^{-1})(|\vec{p}-\vec{q}|^2+\lambda^2)
       (\vec{q}\,^2 +\lambda^2)} ~.
\endeqna
The scattering amplitude due to the KP contact term is 
\eqna
\DE{BSvp}{p} &= &
-4 E_F {(Z\alpha)^2 \over \pi^2 } \int ^1_0 dt \rho_2(t) 
\int_0^{\infty} dp\int_0^{\infty} dq \int^1_{-1}d(cos\theta)
{ \vec{p}\,^2 \over  \vec{p}\,^2 +\lambda^2 }
\nonumber \\ 
&& \times
{ 16  \vec{p}\,^2
 \over (\vec{p}\,^2 +\lambda^2)(\vec{p}\,^2+4m^2(1-t^2)^{-1})
 (\vec{q}\,^2 +\lambda^2)} ~.
\endeqna
The sum of $\DE{BSvp}{n},\DE{BSvp}{o},$and $\DE{BSvp}{p}$
gives the QED contribution $ \Delta E_{\rm QED vp}$ of
(\ref{Eqedvp}).

\subsection{Radiative Photon Contribution}

\begin{figure}
\centerline{\epsfbox{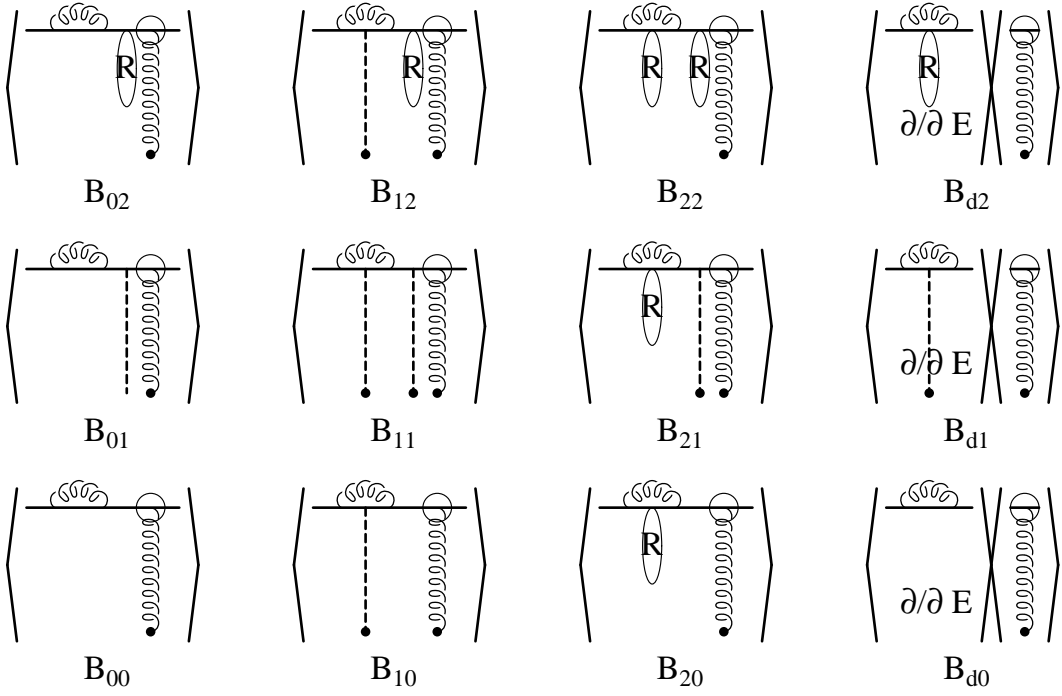}}
\vspace{1ex}
\caption{ Diagrams relevant  to the NRQED bound state contibution $E_B$ of 
the $\alpha(Z\alpha)^2$ correction.
\label{E_Bfig}}
\end{figure}

\begin{figure}
\centerline{\epsfbox{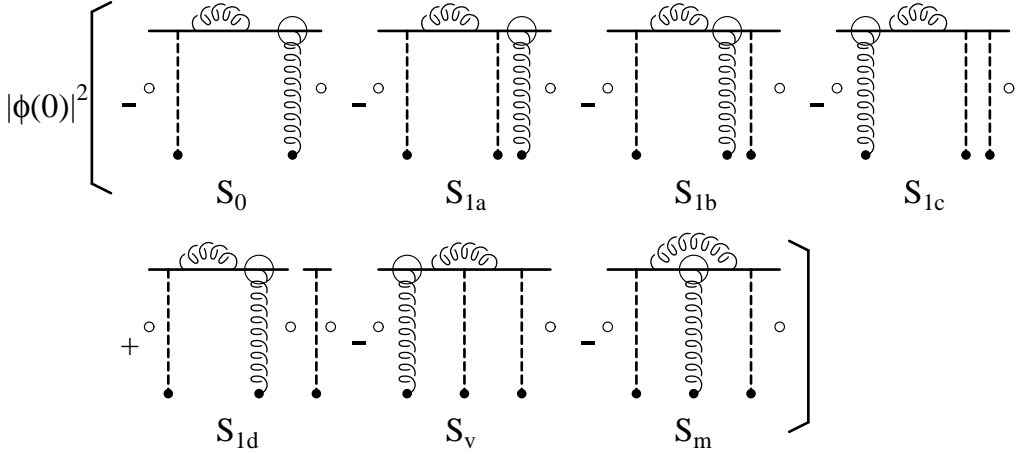}}
\vspace{1ex}
\caption{ Diagrams relevant  to the NRQED scattering state contibution $E_S$ 
of the $\alpha(Z\alpha)^2$ correction.
\label{E_Sfig}}
\end{figure}

%
%
%
%
%
%
We list formulas needed for the calculation of the bound state contribution $E_B$.
For the diagram  $B_{00}$ of Fig. \ref{E_Bfig} we find

\eqna
\DE{E_Bfig}{B_{00}}  & = &
\overall \print \wavef
\pola{p}{p}
\kint   
\fermion{p}{p_0}
\nonumber \\
&=&  E_F {\alpha(Z\alpha)^2 \over \pi }(-{2^3 \over 3}) 
{m\over \gamma} \kint { 2p_0+\gamma \over (p_0+\gamma)^2 }~.
\endeqna
%
%
%
%
%
%
Some contributions of Fig. \ref{E_Bfig} can be similarly 
reduced to integrals over the radiative
photon momentum k while others are harder to simplify:
\eqna
\DE{E_Bfig}{B_{01}}  & = &
 E_F { \alpha(Z\alpha)^2 \over \pi } (-{2^5 \over 3})
     m 
\nonumber \\ & \times &
\kint   \biggl [ -{1\over 4 (p_0^2-\gamma^2 ) } 
    + {p_0^2 \over (p_0^2-\gamma^2)^2} 
   \ln \left({ \gamma+p_0 \over 2\gamma} \right) \biggr ]~,
\\
\DE{E_Bfig}{B_{02}}  &  =  &
 E_F { \alpha(Z\alpha)^2 \over \pi } (-{4 \over 3})
\kint  { m \over (p_0^2 - \gamma^2)^3 (p_0 + \gamma) }
\nonumber \\    & \times  &
\biggr [ 13 p_0^5-27 \gamma p_0^4 + 18 \gamma^2 p_0^3 
- 6 \gamma^3 p_0^2 + \gamma^4 p_0+ \gamma^5
\nonumber \\ & + &
16 \gamma p_0^2 (p_0^2- \gamma^2) 
         \ln \biggl({2 \gamma \over p_0+ \gamma }\biggr)
  \biggr]~,
\\
\DE{E_Bfig}{B_{10}}  & = &
 E_F {\alpha(Z\alpha)^2 \over \pi}{2^5 \over 3} 
m^2 \kint k {1 \over (p_0^2-\gamma^2)^3}
\nonumber \\ && \times
 \left [ {1\over2}(p_0-\gamma)^2
+(3 p_0^2+\gamma^2) \ln( {\gamma + p_0 \over 2\gamma })
- 2 p_0^2 \ln {p_0 \over \gamma}  \right ]~, 
\\
\DE{E_Bfig}{B_{11}}  & = &
\overall \pqlrint \wavef
\pola{p}{q}
\nonumber \\ && \times
\kint k  
\fermion{p}{p_0}
\coulomb{p}{q}
\fermion{q}{p_0}
\nonumber \\ && \times
\fermion{q}{\gamma}
\coulomb{q}{l}
\fermion{l}{\gamma}~,
\\
\DE{E_Bfig}{B_{12}} & = &
\overall \pqlrint \wavef
\nonumber \\ && \times
\vec{p}\cdot\vec{q}
\kint k \fermion{p}{p_0} \coulomb{p}{q} \fermion{q}{p_0}
\nonumber \\ && \times
{ -64 \pi \over (Z\alpha) \gamma^4 } \tilde{R}'(\vec{q},\vec{l};E_0)~,
\\
\DE{E_Bfig}{B_{20}} & = &
\overall \pqrint \wavef
\nonumber \\ && \times
\pola{p}{q}
\kint k 
{ -64 \pi \over (Z\alpha) \gamma^4 } \tilde{R}(\vec{p},\vec{q}; E_0-k)
\fermion{q}{\gamma}~,
\\
\DE{E_Bfig}{B_{21}} & = &
\overall \pqlrint \wavef
\nonumber \\ && \times
\pola{p}{q}
\kint k 
{ -64 \pi \over (Z\alpha) \gamma^4 } \tilde{R}(\vec{p},\vec{q}; E_0-k)
\nonumber \\ && \times 
\fermion{q}{\gamma}
\coulomb{q}{l}
\fermion{l}{\gamma}~,
\\
\DE{E_Bfig}{B_{22}} & = &
\overall \pqlrint \wavef
\nonumber \\ && \times
\pola{p}{q}
\kint k 
{ -64 \pi \over (Z\alpha) \gamma^4 } \tilde{R}(\vec{p},\vec{q}; E_0-k)
{ -64 \pi \over (Z\alpha) \gamma^4 } \tilde{R}'(\vec{q},\vec{l};E_0)~,
\\
\DE{E_Bfig}{B_{d0}} & = &
 E_F  {\alpha(Z\alpha)^2 \over \pi } (-{2^3 \over 3})
m^2 \kint  k  { p_0+5 \gamma \over (p_0+ \gamma)^5 }~, 
\\
\DE{E_Bfig}{B_{d1}} &= &
E_F { 2 \alpha \over 3  \pi }
\pqint 
{( 8\sqrt{\pi\gamma^5})^2 \over  (\vec{p}\,^2 + \gamma^2 ) 
 (\vec{q}\,^2 + \gamma^2 ) }
\pola{p}{q}
\kint k 
\nonumber \\ && \times 
\biggl ({ \partial \over \partial E} 
{ -2m \over \vec{p}\,^2 - 2mE }  
{ -4\pi Z\alpha  \over |\vec{p}-\vec{q}|  }  
{ -2m \over \vec{q}\,^2 - 2mE }  
\biggr )|_{E=-{\gamma^2\over 2m} - k}
\nonumber \\
 &= &
E_F { 2 \alpha \over 3  \pi }
\pqint 
{( 8\sqrt{\pi\gamma^5})^2 \over  (\vec{p}\,^2 + \gamma^2 ) 
 (\vec{q}\,^2 + \gamma^2 ) }
\pola{p}{q} 
\kint k 
\nonumber \\ && \times
 \left [ {2m \over \vec{p}\,^2 +p_0^2 }  +
{ 2m \over \vec{q}\,^2 +p_0^2 }  \right ] 
{ -2m \over \vec{p}\,^2 +p_0^2 }  
{ -4\pi Z\alpha  \over |\vec{p}-\vec{q}|  }  
{ -2m \over \vec{q}\,^2 +p_0^2 } ~, 
\\
\DE{E_Bfig}{B_{d2}} &= &
E_F { 2 \alpha \over 3  \pi }
\pqint 
{( 8\sqrt{\pi\gamma^5})^2 \over  (\vec{p}\,^2 + \gamma^2 ) 
 (\vec{q}\,^2 + \gamma^2 ) }
\pola{p}{q}
\nonumber \\
&& \times \kint k \left ({ \partial \over \partial E} 
{ -64 \pi \over (Z\alpha) \gamma^4 } \tilde{R}(\vec{p},\vec{q}; E)
\right )|_{E=-{\gamma^2\over 2m} - k}~.
\endeqna

For the scattering state, we  put 
\eqn
p_0^2= -2mE = 2m k~.
\endeqn
The NRQED scattering state contributions to $E_S$  
are given by the following terms:
\eqna
\DE{E_Sfig}{S_0} &=&
 -2\overalls \pint \combs{p}\fermis{p} \fermip{p}
\pola{p}{p} \kint \fermion{p}{p_0}
\nonumber \\
&=&  E_F {\alpha(Z\alpha) \over \pi} {8 \over 3} 
 \kint 
{1 \over (p_0^2-\lambda^2)^2 }
 (\lambda^3-3\lambda p_0^2+2p_0^3   ) ~.
\endeqna
Similarly, we find
\eqna
\DE{E_Sfig}{S_{1a}} 
&=&  E_F {\alpha(Z\alpha)^2 \over \pi}{ 32\over 3} 
\kint {1\over p_0^2-\lambda^2}
\ln\left({p_0+2\lambda \over 3\lambda}\right)  ~,
\\
\DE{E_Sfig}{S_{1b}} 
&=&  E_F {\alpha(Z\alpha)^2 \over \pi}{32m \over 3} 
\kint {1 \over p_0^2-\lambda^2 }
  \left [ {p_0\over\lambda}-1 + \logt3 
+ \ln \left( { 2\lambda + p_0 \over \lambda +p_0 } \right)  \right ] ~, 
\\
\DE{E_Sfig}{S_{1c}} 
&=&  E_F {\alpha(Z\alpha)^2 \over \pi}{32m \over 3} 
\kint {1 \over p_0^2- \lambda^2 }
 \left [ \logt3 -{ p_0^2 \over \lambda^2 } 
\ln \left({p_0+\lambda \over p_0+2\lambda}\right)  \right ] ~,
\\
\DE{E_Sfig}{S_{1d}} 
&=&  E_F {\alpha(Z\alpha)^2 \over \pi}{32m \over 3} 
\kint {p_0^2 \over (p_0^2-\lambda^2)^2 } ~,
\\
\DE{E_Sfig}{S_v} 
&=&  -E_F {\alpha(Z\alpha)^2 \over \pi}{32 m^2 \over 3} 
\kintv {1\over (p_0^2-\lambda^2)^2}
\biggl [  \ln \biggl( {(p_0+2\lambda)^2   \over 3 \lambda (2 p_0+\lambda) } 
 \biggr) 
\nonumber \\
&+&(p_0^2-\lambda^2) \biggl\{ {1\over\lambda^2 }
                        \ln\biggl ({ p_0+2\lambda \over p_0 + \lambda} \biggr) 
                           -{1\over p_0^2 }
                       \ln \biggl( { 2 p_0+\lambda \over p_0 + \lambda} \biggr) 
                   \biggr \} 
\biggr ]~,
\\
\DE{E_Sfig}{S_m}
&=&  E_F {\alpha(Z\alpha)^2 \over \pi} {16m^2\over3}
\kintv { 1 \over (p_0^2-\lambda^2)^2} 
 \biggl [
-{(p_0-\lambda)^2 \over p_0^2 } 
\nonumber \\
&+ &
{ (p_0^2-\lambda^2)^2\over p_0^4 }
        \ln \biggl( {p_0+\lambda \over  \lambda }  \biggr) 
 - \ln \biggl( { (p_0+2 \lambda)^2 \over 3 \lambda (p_0+a)}  \biggr)  
\nonumber \\
&+&{ \lambda^2 (2p_0^2-\lambda^2)  \over p_0^4 }  
      \ln \biggl( { 2 p_0 + \lambda \over p_0 + \lambda }  \biggr) 
\biggr ] ~.
\endeqna


The $q^2$ derivative Fermi term $\Delta{\cal{E}}_{q^2}$ contributing 
to $E_R$ is calculated
by replacing $\vec{p}^2 + \vec{p'}\,^2$ 
in $V_W$ of (\ref{Ewa}),  etc. by $ \vec{q}\,^2$.
We obtain
\eqna
\DE{Deri}{a} & = &   \int\frac{d^{3}p}{(2\pi)^3}
                       \int^{\Lambda}\frac{d^{3}k}{(2\pi)^3}
              \frac{(8\sqrt{\pi\gamma^{5}})^{2}}
              {(|\vec{p}-\vec{k}|^{2}+\gamma^{2})^{2}
               (\vec{k}\,^2+\gamma^{2})^{2}}
          \langle V_{\vec{q}\,^2}(\vec{k})  \rangle
\nonumber \\
   & = & E_{F}(Z\alpha)^{2}
       \left [-2\frac{\Lambda}{\gamma\pi} +3\right ] ~,
\label{Eq2a}
\\
\DE{Deri}{b} & = &
          E_{F}(Z\alpha)^{2} 2\frac{\Lambda}{\gamma \pi} ~,
\\
\DE{Deri}{c} & = &
          E_{F}(Z\alpha)^{2}
          \left [4\frac{\Lambda}{\lambda \pi} +2\ln\frac{2}{3} -3 \right ] ~,
\\
\DE{Deri}{d} & = &
          E_{F}(Z\alpha)^{2}
  \left [4\frac{\Lambda}{\lambda \pi} +2\ln2 -\ln 3 -{9\over2} \right ] ~,
\\
\DE{Deri}{e} & = &
   E_{F}(Z\alpha)^2   \left [-8 {\Lambda \over \lambda\pi} + 6   \right ]~.
\label{Eq2z}
\endeqna
The sum of the contributions (\ref{Eq2a}) --(\ref{Eq2z}) is
\eqn
\Delta E_{q^2}=  E_{F}(Z\alpha)^{2}
 \left [ 4\ln 2-3\ln 3  +\frac{3}{2}  \right ]~.    \label{Eq2}
\endeqn
Multiplying  $\Delta E_{q^2}$ with 
the $``$renormalization" constant $c_{q^2}$ of (\ref{Rconsts}), we obtain
the contribution $\Delta {\cal{E}}_{q^2}$ listed in Table \ref{E_Rtable}.

\begin{figure}
\centerline{\epsfbox{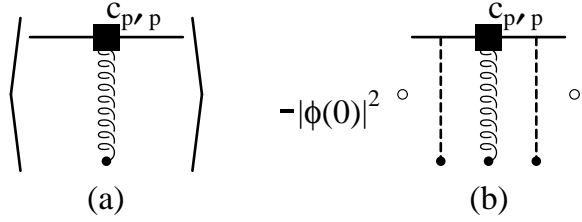}}
\vspace{1ex}
\caption{ Diagrams relevant  to the $``$renormalization" constant
contibution $E_R$ of the $\alpha(Z\alpha)^2$ correction.
Only the diagrams involving $p'p$ interaction term are shown here.
For other terms, see Figs. 2,3,4, and 5.  
\label{E_Rfig}}
\end{figure}

The reminder is the contribution $\Delta {\cal{E}}_{p'p}$ coming from
the $p'p$ coupling interaction.  The $p'p$ potential $V_{p'p}(\vec{p'},
\vec{p}) $ for spherical symmetric states is defined by
\eqn
V_{p'p}(\vec{p}\,',\vec{p})= { Z\alpha \pi \over 6 m^3 M } 
{ -\vec{p}\,^2 \vec{q}\,^2 + (\vec{p}\cdot \vec{q})^2  
\over \vec{q}\,^2 + \lambda^2 } \vec{\sigma}_e \cdot \vec{\sigma}_{\mu}~.
\label{Vpp}
\endeqn
With this expression, the bound state calculation of Fig. \ref{E_Rfig}(a)
gives
\eqna
\DE{E_Rfig}{a} & = & \int\frac{d^{3}p}{(2\pi)^3}
                       \int^{\Lambda}\frac{d^{3}k}{(2\pi)^3}
              \frac{(8\sqrt{\pi\gamma^{5}})^{2}}
              {(|\vec{p}-\vec{k}|^{2}+\gamma^{2})^{2}
               (\vec{k}\,^2+\gamma^{2})^{2}}
          \langle V_{p'p}(\vec{p}-\vec{k},\vec{p})  \rangle
\nonumber \\
   & = & E_{F}(Z\alpha)^{2}
     \biggl [-{1 \over 2} \ln \biggl (\frac{\Lambda}{\gamma} \biggr )
        +{1\over 2} + {1 \over 2} \ln 2 \biggr ] ~.
\label{Eppa}
\endeqna
The 2-loop scattering diagram of Fig. \ref{E_Rfig}(b) gives  the contribution 
\eqna
\DE{E_Rfig}{b} & = & \int\frac{d^{3}p}{(2\pi)^3}
                       \int^{\Lambda}\frac{d^{3}k}{(2\pi)^3}
V_C(0,\vec{p}-\vec{k}) {-2m \over |\vec{p}-\vec{k}|^2 }
          \langle V_{p'p}(\vec{p}-\vec{k},\vec{p})  \rangle
{-2m \over \vec{p}\,^2  } V_C(\vec{p},0) 
\nonumber \\
   & = & E_{F}(Z\alpha)^{2}
     \biggl [{1 \over 2} \ln \biggl (\frac{\Lambda}{\lambda} \biggr )
        +{1\over 8} - {3 \over 4} \ln 3 \biggr ] ~.
\label{Eppz}
\endeqna
The sum of (\ref{Eppa}) and (\ref{Eppz})  is
\eqn
\Delta E_{p'p} =  E_{F}(Z\alpha)^{2}
     \biggl [-{1 \over 2} \ln \biggl (\frac{\lambda}{\gamma} \biggr )
        +{5\over 8} + {1 \over 2} \ln 2 -{3 \over 4} \ln 3 \biggr ] ~,
\endeqn
where $\lambda$ is the infrared cutoff.
Multiplying this with the
$``$ renormalization" constant $c_{p'p}=a_e$ of (\ref{Rconsts}), we find 
the contribution $\Delta {\cal{E}}_{p'p}$ listed in Table III.

\begin{figure}
\centerline{\epsfbox{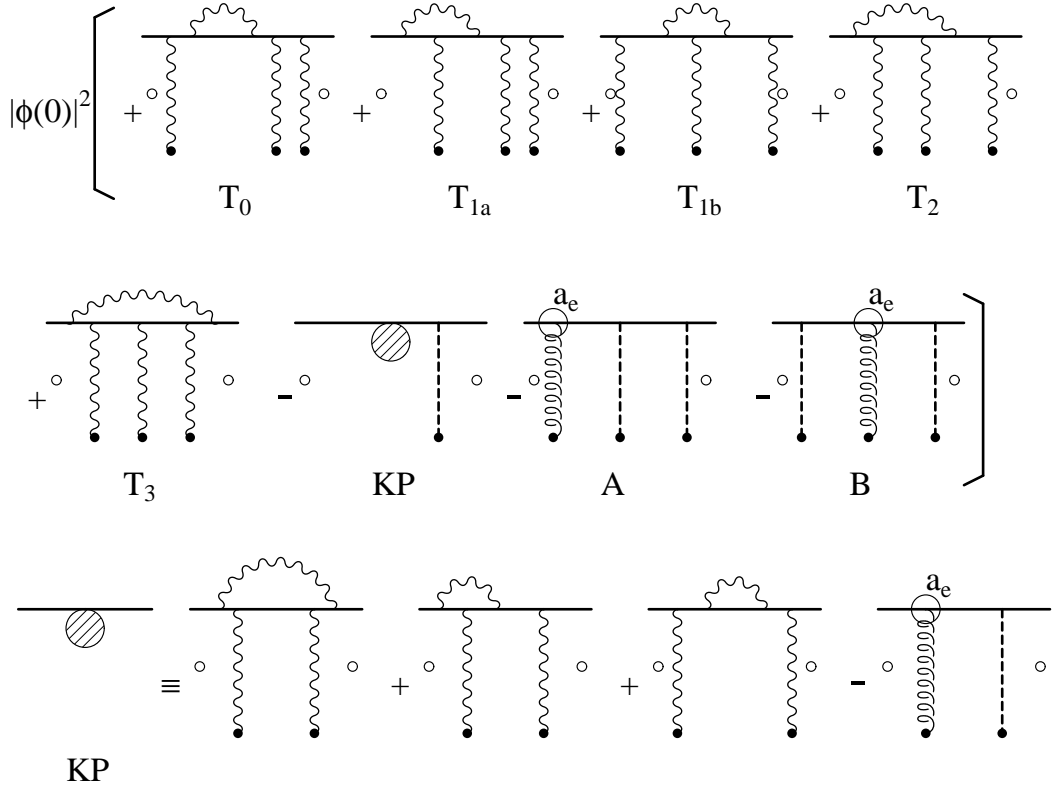}}
\vspace{1ex}
\caption{ Diagrams relevant  to the QED scattering state contibution $E_Q$ 
of the $\alpha(Z\alpha)^2$ correction.
The last line is the difinition of the KP potential which contribute
to the $\alpha(Z\alpha)$ correction of the hyperfine splitting.
\label{E_Qfig}}
\end{figure}

The contribution $E_Q$ of the QED scattering diagrams shown in Fig. \ref{E_Qfig}
can be calculated using the techniques similar to one described 
in Ref.\cite{KN1}. These diagrams may be  expressed in the form:
\eqna
2 ie^2 (Ze^2)^3{\gamma^3 \over \pi}
\int {d^{4}k \over (2\pi)^4}
\int {d^{4}p \over (2\pi)^4}
\int {d^{4}q \over (2\pi)^4} 
{
{\cal{E}}_{\mu \nu \lambda} 
{\cal{M}}^{\mu \nu \lambda}
\over 
 (p^2 - \lambda^2 )
((p-q)^2 - \lambda^2 ) (q^2 - \lambda^2 ) } 
\endeqna
where ${\cal{E}}_{\mu \nu \lambda}$ and $ {\cal{M}}^{\mu \nu \lambda}$
are factors representing the electron line  and the muon line, respectively. 
For instance,
${\cal{E}}_{\mu \nu \lambda}$ corresponding to the diagram 
$T_{1a}$ of Fig.\ref{E_Qfig} is given by 
\eqn
{\gamma_{\mu}( \not p +\not l + m ) \gamma_{\nu} ( \not q +\not l + m ) 
\gamma^{\alpha}( \not k + \not q +\not l + m ) \gamma_{\lambda} 
(\not k +\not l + m ) \gamma_{\alpha}
\over 
((p+l)^2-m^2+i\epsilon)
((q+l)^2-m^2+i\epsilon)
((k+q+l)^2-m^2+i\epsilon)
((k+l)^2-m^2+i\epsilon)   } ~,
\endeqn
where $l=(m,\vec{0})$.   ${\cal{M}}^{\mu \nu \lambda}$ may 
be written as the sum of six permutation terms:
\eqn
{\gamma^{\mu}( - \not p +\not r + M ) \gamma^{\nu} ( -\not q  +\not r + M ) 
  \gamma^{\lambda} 
\over 
((-p+r)^2-M^2+i\epsilon) ((-q+r)^2-M^2+i\epsilon) }
+ {\rm permutations} ~~{\rm in}~~ \mu,~~\nu,~~\lambda ~,
\endeqn
where $r=(M,\vec{0})$.

%
The integral is greatly simplified  in the limit of  infinite muon
mass.  We can extract the contribution to the hyperfine splitting from
each diagram using the projection operator given by Eq. (65) of \cite{KN1}.
For instance, the contribution from the diagram $T_{1a}$ of Fig.\ref{E_Qfig},
after carrying out the $k$ integration and subtracting 
the vertex renormalization term,
is expressed  with the help of Feynman parameters $z_1$, $z_2$, and $z_5$ as   
\eqna
\DE{E_Qfig}{T_{1a}} &= & 2 E_F {\alpha(Z\alpha)^2 \over \pi}
{1 \over 8\pi^4}
\int dz_1 dz_2 dz_5 \delta(1-z_1-z_2-z_5)
\nonumber \\ &&
\int d^{3}p 
\int d^{3}q 
{ 1 \over \vec{p}\,^2 \vec{q}\,^2 
          (\vec{p}\,^2 + \lambda^2 )
          (|\vec{p}-\vec{q}|\,^2 + \lambda^2 )
          (\vec{q}\,^2 + \lambda^2 )   }
\nonumber \\ &&
\times \biggl[
-8 \ln \biggl ( {V \over V_0} \biggr )
     ( \vec{p}\,^2 - \vec{p}\cdot \vec{q} + \vec{q}\,^2 ) 
\nonumber \\ &&
+(\vec{p}\,^2-\vec{p}\cdot\vec{q} ) (8+8A_1^2-32 A_1)( {1\over V}-{1\over V_0})
\nonumber \\ &&
+\vec{q}\,^2( 8 - 24A_1){1\over V}
-\vec{q}\,^2(8+8A_1^2-32A_1){1\over V_0}
\nonumber \\ &&
+{\vec{q}\,^2 \over V} \biggl \{  \vec{p}\,^2 
            ( -4 A_1~ A_{1q}-4 A_1~A_{2q} + 8 A_{1q} A_{2q} )
\nonumber \\ &&
+ \vec{p}\cdot\vec{q}
                    ( 4 A_1~ A_{2q} - 8 A_{1q} A_{2q} )
+\vec{q}\,^2  ( 8 A_{1q}~ A_{2q})  \biggr \}
\biggr ],
\label{T1a}
\endeqna
where
\eqna
&&
A_1=z_5,~~~~~A_{1q}=1-z_1,~~~~~A_{2q}=-z_1,
\nonumber \\&&
V= z_1+z_2-(z_1+z_2)A_1 + z_1 ~A_{1q} \vec{q}\,^2 ,
\nonumber \\&& 
V_0= z_1+z_2-(z_1+z_2)A_1 . 
\endeqna
The integral $\DE{E_Qfig}{T_{1a}}$ has  one threshold singularity 
at $\vec{q}=0$ and another at $\vec{p}=0$.
The threshold singularity  at $\vec{q}=0$ is canceled by 
that of the NRQED scattering diagram $\DE{E_Qfig}{A}$ which consists of
one Fermi and two Coulomb potentials multiplied by the Fermi-term 
$``$ renormalization" constant of the Fermi term $a_e$, 
namely  the anomalous magnetic moment
of the electron.  The latter is of the form 
\eqna
\DE{E_Qfig}{A} &=&  -2 E_F {\alpha(Z\alpha)^2 \over \pi}
{1 \over 8\pi^4}
\int dz_1 dz_2 dz_5 \delta(1-z_1-z_2-z_5)
\nonumber \\  \times 
\int d^{3}p 
\int d^{3}q & & 
{ 1 \over \vec{p}\,^2 \vec{q}\,^2 
          (\vec{p}\,^2 + \lambda^2 )
          (|\vec{p}-\vec{q}|\,^2 + \lambda^2 )
          (\vec{q}\,^2 + \lambda^2 )   }
\vec{q}\,^2 (-8A_1^2+8A_1){1 \over V_0} ~~.  
\label{A1}
\endeqna
The threshold singularity at $\vec{p}=0$ is 
canceled by the contact term consisting of  
the lower order contact term (the KP term),
which contributes to the $\alpha(Z\alpha)E_F$  correction, and the Coulomb 
potential.   
>From  the second and fourth diagrams of the KP diagram of  
Fig. \ref{E_Qfig}(KP)  we find  that 
the contact term contribution corresponding to 
$T_{1a}$ and $A$ of Fig. \ref{E_Qfig} is given by 
\eqna
 & &  -2 E_F {\alpha(Z\alpha)^2 \over \pi}
{1 \over 8\pi^4}
\int dz_1 dz_2 dz_5 \delta(1-z_1-z_2-z_5)
\nonumber \\ & &
\int d^{3}p 
\int d^{3}q 
{ 1 \over \vec{p}\,^2 \vec{q}\,^2 
          (\vec{p}\,^2 + \lambda^2 )
          (\vec{q}\,^2 + \lambda^2 )^2   }
\nonumber \\ & &
\times\biggl[
-8 \ln \biggl ( {V \over V_0} \biggr )
     \vec{q}\,^2  
+\vec{q}\,^2( 8 - 24A_1)({1\over V}-{1\over V_0} )
+(\vec{q}\,^2)^2  ( 8 A_{1q}~ A_{2q}) {1\over V} 
\biggr ] ~.
\label{KPt1aa1}
\endeqna
The sum of (\ref{T1a}), (\ref{A1}),  and (\ref{KPt1aa1}) still
suffers from a severe infrared singularity 
in the limit of vanishing radiative photon mass. 
In order to perform numerical integration we identified the IR singular
terms of (\ref{T1a}) and (\ref{KPt1aa1}) and subtracted them
>from each integral.
The IR subtraction term for  (\ref{T1a}) is
of the form:
\eqna
\DE{E_Qfig}{T_{1a}}_{IR} &= & -2 E_F {\alpha(Z\alpha)^2 \over \pi}
{1 \over 8\pi^4}
\int dz_1 dz_2 dz_5 \delta(1-z_1-z_2-z_5)
\nonumber \\ &&
\int d^{3}p 
\int d^{3}q 
{ 1 \over \vec{p}\,^2 \vec{q}\,^2 
          (\vec{p}\,^2 + \lambda^2 )
          (|\vec{p}-\vec{q}|\,^2 + \lambda^2 )
          (\vec{q}\,^2 + \lambda^2 )   }
\nonumber \\ &&
\times 
(\vec{p}\,^2-\vec{p}\cdot\vec{q} + \vec{q}\,^2) (-16)( {1\over V_{IR}}
-{1\over V_0 } )~~,
\label{T1aIR}
\endeqna
where
\eqn
V_{IR}= z_1+z_2-(z_1+z_2)A_1 + z_1 \vec{q}\,^2  ~~. 
\endeqn
This IR subtraction term  is completle canceled by
that  for  $T_{0}$ of Fig.  \ref{E_Qfig}.
Similar cancellation occurs among the diagrams $T_{1b}$,
$T_{2}$, and $T_3$. 
The IR subtraction term for Eq. (\ref{KPt1aa1})  is given by 
\eqna
& &  +2 E_F {\alpha(Z\alpha)^2 \over \pi}
{1 \over 8\pi^4}
\int dz_1 dz_2 dz_5 \delta(1-z_1-z_2-z_5)
\nonumber \\ & &
\int d^{3}p 
\int d^{3}q 
{ 1 \over \vec{p}\,^2 \vec{q}\,^2 
          (\vec{p}\,^2 + \lambda^2 )
          (\vec{q}\,^2 + \lambda^2 )^2   }
\nonumber \\ & &
\times  
\vec{q}\,^2( -16)({1\over V_{IR}}-{1\over V_0} ) 
\biggr ] ~.
\endeqna
This type of IR sungularities of the KP contact terms cancel out completely
among themselves.
When summed over
all diagrams of Fig. \ref{E_Qfig},  the resultant integrand has 
only the 
infrared singular terms  of the form $\ln^2 (\lambda/m)$  
and $\ln (\lambda/m)$.


\vspace{2cm}

\vspace{2cm}

\vspace{2cm}

\vspace{5cm}

\vspace{2cm}

\begin{table} 
\caption{Bound state contibutions $E_B$ of the $\alpha(Z\alpha)^2$
correction from individual diagrams shown  in Fig. 7.
The overall factor $E_F \alpha(Z\alpha)^2 / \pi$ is omitted.
\label{E_Btable}
}
\[
\begin{array}{ccl}
\hline \hline
{\rm Diagram} & & ~~~~~~~~~~~~{\rm Contribution}
\\ \hline
B_{00} & &-{ 16 \over 3 } \biggl [ \sqrt{{2\Lambda m \over \gamma^2} }
                      - { 3 \over 4} \logtLg 
              +{3 \over 2} \ln 2 - {3 \over 4}   \biggr]  
\\ \hline
B_{01} & &-{ 16 \over 3 }    \biggl [  { 1 \over 4} \logstLg  
                    +\biggl (- \ln 2 - {1 \over 4} \biggr )\logtLg  
+{1\over 2} \ln 2 + \ln^2 2  + {\pi^2 \over 6}    \biggr ]
\\ \hline
B_{02} & &-{ 16 \over 3 } \biggl [ {13 \over 8} \logtLg 
-{49 \over 12}-{13 \over 4} \ln 2 - {\pi^2 \over 8 }    \biggr ]
\\ \hline
B_{10} & & { 8 \over 3 }    \biggl [  { 1 \over 4} \logstLg 
                    +\biggl( -3 \ln 2 +  {1 \over 2} \biggr )\logtLg 
            +\ln 2 +3 \ln^2 2   +{\pi^2 \over 3 } - {3\over 2}    \biggr ]
\\ \hline
B_{11} & & { 8 \over 3 } \biggl( { \pi^2 \over 6 } -  {1 \over 2} \biggr) 
\biggr(\logtLg -1 \biggl)
-10.499~478~(61)
\\ \hline
B_{12} & & { 8 \over 3 } \biggl (-{ \pi^2 \over 6 } +  {9 \over 4} \biggr)
\biggl (\logtLg -1 \biggr )
-5.393~042~(55)
\\ \hline
B_{20} & &+0.761~648~(52)
\\ \hline
B_{21} & &+0.491~192~(42)
\\ \hline
B_{22} & &+0.221~861~(36)
\\ \hline
B_{d0} & & -{ 8 \over 3 } \biggl [{1 \over 4} \logtLg -{1\over2}\ln2 
- {5 \over 24} \biggr ] 
\\ \hline
B_{d1} & &-0.316~379~(13)
\\ \hline
B_{d2} & &-0.116~135~(19)
\\ \hline \hline 
\end{array}
\]
\end{table}

\begin{table}  
\caption{Scattering state contributions $E_S$ of the $\alpha(Z\alpha)^2$
correction from individual diagrams shown  in Fig. 8.
The overall factor $E_F \alpha(Z\alpha)^2 / \pi$ is omitted.
$Li(1/3)=0.366~213~229~\cdots $ is the value of dilogarithmic 
function $Li(x)$ at $x=1/3$, 
where $Li(x)=-\int^x_0 dt {\ln(1-t) \over t }$. 
\label{E_Stable}
}
\[
\begin{array}{ccl}
\hline \hline
{\rm Diagram }&    &~~~~~~~~~~~~  {\rm Contribution }
\\ \hline
S_0 & &{ 16 \over 3 } \sqrt{{2\Lambda m \over \gamma^2} }
\\ \hline
S_{1a} & &{ 16 \over 3 }    \biggl [  { 1 \over 4} \logstLl 
                     - \ln 3 \logtLml 
    + {1\over2}\ln^2 3 + {\pi^2 \over 4 } + Li({1\over3})   \biggr ]
\\ \hline
S_{1b} & &{ 16 \over 3 }  \biggl [ 2 \sqrt{{2\Lambda m \over \lambda^2} }
+\biggl (\logt3 -1 \biggr )\logtLml
    + {1 \over 2} \ln^2 3 + Li({1 \over 3})   \biggr ]
\\ \hline
S_{1c} & &{ 16 \over 3 }  \biggl [ 2 \sqrt{{2\Lambda m \over \lambda^2} }
+\biggl(\logt3 -{3\over2} \biggr)\logtLml
    + {1 \over 2 } \ln^2 3 + 4 \ln 2 - {3 \over 2} + Li({1 \over 3}) \biggr ]
\\ \hline
S_{1d} & &{ 16 \over 3 }  \biggl [ -4 \sqrt{{2\Lambda m \over \lambda^2} }
+3\logtLml -2 \biggr ] 
\\ \hline
S_v & & -{ 8 \over 3 }    \biggl [  { 1 \over 4} \logstLl 
                    + \biggl(  -2\ln2 - \ln 3 +{1\over 2}\biggr) \logtLml
\\
    & &~~ + {\pi^2 \over 4 } +  5 Li({1 \over 3}) 
          -2 \ln^2 2 + {5 \over 2} \ln^2 3  
          + {8 \over 3} \ln 2 + \ln 3 - {7 \over 2}     \biggr ]
\\ \hline
S_m & & -{ 8\over 3 } \biggl [ \biggl( -{1\over2} \ln3 
                   +{1\over2} \biggr)\logtLml
       - { \pi^2 \over 24 } + {3 \over 2} Li({1 \over 3}) 
           + {3 \over 4} \ln^2 3  
          - {2 \over 3} \ln 2 + {1 \over 2} \ln 3 - 1     \biggr ]
\\ \hline \hline 
\end{array}
\]
\end{table}

\begin{table}
\caption{Renormalization constant  contributions $E_R$ of 
the $\alpha(Z\alpha)^2$
correction from individual diagrams shown  in Fig. 9.
The overall factor $E_F \alpha(Z\alpha)^2 / \pi$ is omitted.
\label{E_Rtable}
} 
\[
\begin{array}{ccl}
\hline \hline
{\rm Diagram }&    &~~~~~~~~~~~~  {\rm Contribution }
\\ \hline
\Delta {\cal{E}}_d & & \biggl[3 \logt3 - \loglg \biggr] 
          \biggl [ {8 \over 3}\logm2L+{11\over 9} + 2 a_e + a_e   \biggr ]
\\ \hline
\Delta {\cal{E}}_k & &  \biggl[-{1\over2}-6 \logt3 +2 \loglg \biggr] a_e 
\\ \hline
\Delta {\cal{E}}_s & & \biggl[-{1\over2}-\logt3 - \loglg \biggr] 2 a_e 
\\ \hline
\Delta {\cal{E}}_{q^2} & &  \biggl [ {3\over2}+ 4\ln2 - 3\ln3 \biggr ]  
 \biggl [ {4 \over 3}\logm2L+{11\over 18} 
+ {1\over2 }a_e + {1\over3}   \biggr ]
\\ \hline
\Delta {\cal{E}}_{p'p} & & \biggl[{5\over 8}+ {1\over2}\ln2 
         - {3\over 4} \ln3 -{1\over2}\loglg \biggr] a_e 
\\ \hline \hline
\end{array}
\]
\end{table}

\begin{table}   
\caption{ The QED scattering state contribution $E_Q$ for
various photon mass $\lambda$.  The corresponding diagrams are shown in  
Fig. 10. 
The overall factor $E_F \alpha(Z\alpha)^2 / \pi$ is omitted.
The  uncertainty  of $E_Q$ is from numerical  integration
by VEGAS.
\label{E_Qtable}
}
\vspace{3ex}
\[
\begin{array}{ccccccc}
\hline \hline 
\lambda^2/m^2 & &   E_Q  & &  E_Q -(E_Q)_{\rm log}  & &
{\rm uncertainty}~~{\rm in}~~E_Q
\\ \hline
~~~10^{-5}        & ~~~& -90.075~1 & &  -8.697~1 & & 0.0024 
\\ \hline
~~~10^{-5.5}      & ~~~& -107.943~8 & &  -8.707~8 & & 0.0033 
\\ \hline
~~~10^{-6}        & ~~~& -127.575~1 & &  -8.713~9 & & 0.0040       
\\ \hline
~~~10^{-6.5}      & ~~~& -148.971~3 & &  -8.717~6 & & 0.0048 
\\ \hline
~~~10^{-7}        & ~~~& -172.139~0 & &  -8.725~5 & & 0.0029 
\\ \hline
~~~10^{-7.5}      & ~~~& -197.062~7 & &  -8.722~0 & & 0.0030 
\\ \hline
~~~10^{-8}        & ~~~& -223.758~9 & &  -8.723~7 & & 0.0026 
\\ \hline
~~~10^{-9}        & ~~~& -282.449~4 & &  -8.723~4 & &  0.0016 
\\ \hline \hline 
\end{array}
\]
\label{E_Qdata}
\end{table}

\end{document}